The network limits of infectious disease control via occupation-based targeting

Demetris Avraam[1,2], Nick Obradovich[1], Niccolò Pescetelli[1], Manuel Cebrian[1+] & Alex Rutherford[1+]

[1] Centre for Humans and Machines, Max Planck Institute for Human Development
[2] Population Health Sciences Institute, Newcastle University

[+] Corresponding authors: Alex Rutherford (rutherford@mpib-berlin.mpg.de) & Manuel Cebrian (cebrian@mpib-berlin.mpg.de).

# Abstract

Policymakers commonly employ non-pharmaceutical interventions to manage the scale and severity of pandemics. Of non-pharmaceutical interventions, social distancing policies -- designed to reduce person-to-person pathogenic spread -- have risen to recent prominence. In particular, stay-at-home policies of the sort widely implemented around the globe in response to the COVID-19 pandemic have proven to be markedly effective at slowing pandemic growth. However, such blunt policy instruments, while effective, produce numerous unintended consequences, including potentially dramatic reductions in economic productivity. Here we develop methods to investigate the potential to simultaneously contain pandemic spread while also minimizing economic disruptions. We do so by incorporating both occupational and network information contained within an urban environment, information that is commonly excluded from typical pandemic control policy design. The results of our method suggest that large gains in both economic productivity and pandemic control might be had by the incorporation and consideration of simple-to-measure characteristics of the occupational contact network. However we find evidence that more sophisticated, and more privacy invasive, measures of this network do not drastically increase performance.

# Introduction

Containment of infectious disease requires implementation of strategies to reduce the spread of the disease. Without widespread availability of a robust vaccine, as is likely to be the case for the COVID-19 epidemic in the immediate future, the leading Non-Pharmaceutical Intervention (NPI) is social distancing including closure of businesses[1]: a reduction of the number of people in physical proximity.

Since the European Industrial Revolution workers have increasingly migrated to towns and cities to convene in centralised workplaces.[2] A large literature of work relates growth, creativity and other metrics of economic performance directly to the increasing returns and economies of scale made possible by the density and proximity of workers in regions, cities and workplaces.[3,4,5] Therefore, social distancing is unavoidably disruptive to these workers, workplaces and economies. The highly infectious nature of COVID-19 along with the disease's long incubation period[6,7] have shown the cost of such disruption even more starkly than previous epidemics. As a result, working hour losses globally could be as high as 12% in the third quarter of 2020[8] and a 6·6% drop in global GDP in 2020.[9] For this among other reasons, scholars have noted that the COVID-19 epidemic has the potential to affect the future of work substantially over the long term.[10,11]

More precisely, social distancing policies affect *workplaces* directly and workers indirectly. While some work environments are conducive to social distancing (e.g. construction sites), others are inherently not (e.g. gyms & nightclubs).[12] As a result, social distancing requirements have produced heterogeneous effects on workers that are primarily determined by workers' ability to socially distance in their workplace and/or to work from home (WFH).[13]

Thus policymakers have the difficult task of moving beyond coarse economic interventions e.g. furloughing all workers, to more targeted interventions that balance the human toll of infection with the economic cost of interventions. Stated precisely, workers should be partitioned into groups who may respectively (i) remain working in their workplaces (ii) work effectively from home or (iii) be furloughed, such that the spread of the epidemic is minimised while also minimising economic disruption associated with the intervention.

This optimization task is currently impossible on a practical level for many reasons. These include the lack of sufficiently detailed information on human mobility and social interactions, the complex network dynamics of disease spread,[14] the interconnected nature of modern economies,[15,16,17] and divergent opinions on the proper balance between human and economic loss.[18]

Epidemics typically pose differing levels of risk to individuals based on demographic categories, such as age groups. This may be due to characteristics of the specific disease or due to differing physical contact patterns encoded in contact matrices.[19] This heterogeneity in the spread between members of a population also contributes to the overall complexity of accurately modeling the disease spread.

It is notoriously difficult to measure, at scale, physical proximity as it relates to airborne or droplet mediated transmission of infectious disease.[20,21] Deriving a *network* of contacts is even more challenging compared to a simple count of the number of contacts. This is due to privacy concerns and limitations of measurement accuracy as individuals need to be uniquely identified throughout the measurement.[22] Typically contact interactions are measured through self-reported surveys and networks measured using sensors.[23,24,25,26,27] Previous work has constructed contact matrices between subpopulations stratified by age and other demographics.[28,29] Full contact networks have been measured in large scale field studies in schools,[30,31] dormitories,[32] hospitals,[33] and conferences[34] as well as being approximated from the use of location-based services.[35]

Yet, demographics represent only one set of individual characteristics that likely play a role in modulating differential epidemic spread. An individual's work -- their occupation -- encapsulates many determinants of their epidemic risk. Commuting patterns, the extent of contact with others at their work, and the ease with which non-pharmaceutical interventions such as wearing of personal protective equipment and remote working can be adopted are all occupation-specific factors that shape both individual risk and the propensity for individuals in any given occupation to shape epidemic spread on human networks more generally.

While studies have measured contact networks in workplaces,[36] to date we are not aware of any in-depth study on contact networks stratified by occupation, and as such the true structure of such a network is unknown.

In this work, we examine the role of occupation in an epidemic spread in an urban environment. We also evaluate the efficacy of occupation-based disease control measures within a simulation of epidemic dynamics using the Susceptible-Exposed-Infected-Recovered (SEIR) framework. More specifically, we investigate the effects of several network-based interventions and compare them to the outcomes produced by more coarse heuristics. A focus on occupational interventions, coupled with detailed data on the distribution of the workforce across occupations, wage and workplace proximity allows us to simulate the economic impact of particular containment strategies alongside each intervention's epidemiological impact.

Our methods enable us to approximate the degree of physical contact between individuals within and across occupations without explicit contact matrices (such matrices to our knowledge do not exist across the full empirical network due to the cost and methodological complexity of measuring such contacts). Our approach is general and applicable to any infectious disease, however, we parameterise our simulations with estimations of the epidemiological characteristics of COVID-19 for illustrative purposes.

Our work complements a rapidly growing body of work analysing the effects of social distancing and furloughing policies on human mobility and behaviour in the COVID-19 epidemic[37,38,39,40,41,42,43,44] with several focusing on the economic aspects in particular.[45,46]

Using these methods, we make three marked contributions to the science of epidemic control. First, we contribute a technique for constructing occupation-based epidemic simulations using publicly available data. Second, we compare several occupation-based containment policies based on their epidemiological and economic costs. Third, having identified contact degree as an effective heuristic for containment, we extend to investigate the fundamental limits of the controllability of epidemics in networks.

# Data & Methods

We present a general method for simulating occupation-based epidemic control policy interventions. However, we consider data from New York City as a paradigmatic example.[47] We make use of the Occupational Information Network (O*NET) which is a public repository of occupational data in the US and statistical economic information collected by the Bureau of Labour Statistics (BLS).

## O*NET: occupations and work characteristics

We make use of the O*NET data on "work characteristics" to derive a composite measure of proximity from five distinct work characteristics that are likely to be correlated with the degree of in-person contact a worker is required to have. We inspect the occupations that are assigned high and low degree to manually validate this method (see SI). The projection on the first principal component (PC) is listed below for the top/bottom five jobs.

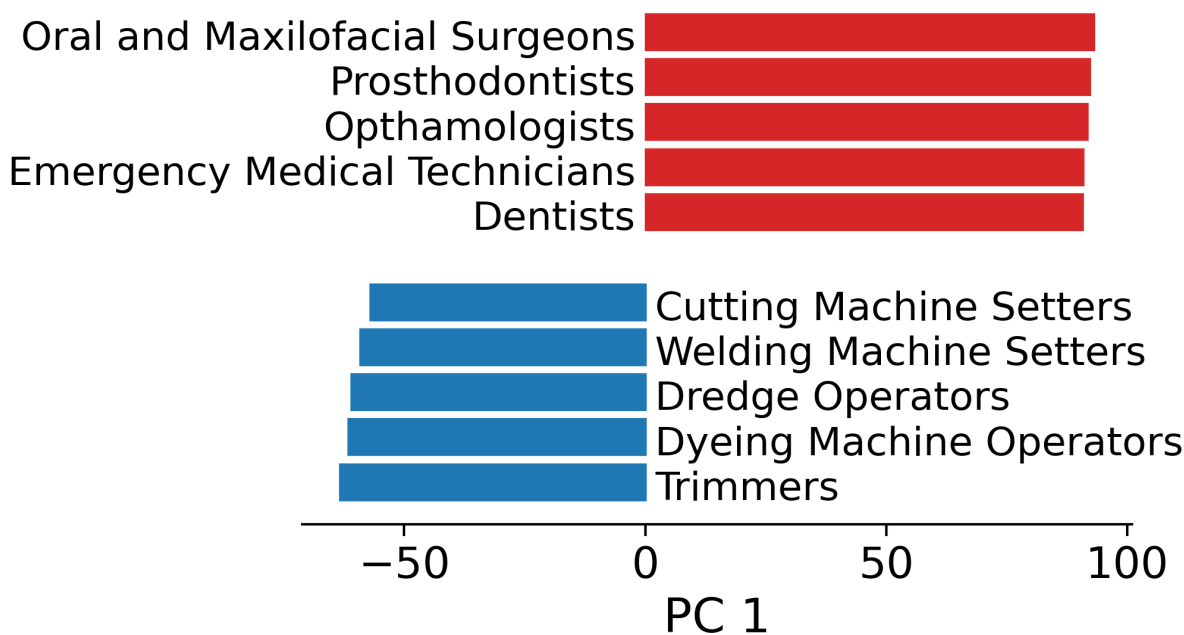

**Figure 1: The five occupations with largest and smallest projection onto the first principal component of proximity measures. The PCA operates on five independent measures of proximity required by each job taken from the O*NET database:** *Exposed to disease or infections, Performing for or working directly with the public, Communicating with persons outside of the organisation, Deal with external customers, Physical proximity.* **The first Principal Component can explain 53% of the total variance. See SI for more details.**

## Contact data: proportions of home/work/transit

In agreement with previous empirical studies of contact networks,[48] we define three categories of contacts among the adult working population: home, work and transport. The mean total contact degree has been found to be 75 in a recent study of proximity via Bluetooth[37] in New York City. Fixing the mean home degree at two based on the mean household size,[49] and considering an equal work and transportation mean degree, we set the ratio of the mean contact degree to be 2:36·5:36·5 for home:work:transport, respectively (see SI). We fit the empirical data found in[29] to derive a log-normal distribution for each category rescaling the mean to fit the above ratios. For each worker node, we sample each of the three distributions to determine total degree. The number of work contacts for each worker depends on her occupation, while the number of home and transportation links are independent of occupation. Consider a worker in a high proximity occupation such as a Retail Salesperson, links are assigned as follows:

1. The proximity score of the occupation is taken as the projection of the job on the first principal component derived from the PCA of the proximity variables. For the case of a retail Salesperson this is 0·55 in the range [0-1]
2. This specific PCA value is mapped to a log-normal distribution (mean of 36·5) of work degree by mapping the percentiles of the distributions yielding a work degree of 49.
3. A transport degree is assigned to each worker node individually by independently sampling a log-normal distribution with a mean of 36·5.
4. A home degree is assigned to the worker by independently sampling a log-normal distribution with a mean of 2, yielding a value of 2.

Links from each category are functionally equivalent in terms of disease transmission in our simulations. However each layer is independent, and different policy interventions retain or remove links from each layer as described later.

## Work from Home data

Some workers can work from home without disruption to productivity and thus may be effectively removed from the occupational contact network without loss of economic contribution. We consider a binary index for O*NET jobs derived by Dingel and Neiman.[13]

## Essentialness data

Since one motivation for NPI selectively furloughs workers according to perceptions of the essentialness of their contributions, we assign a measure of 'essentialness' using the data of del Rio-Chanona et al (2020).[45,50]

## Wages data

Another NPI we consider here to balance between the loss of economic productivity and the epidemic expansion is the furlough of workers prioritising those on a low income. We use wage data by occupation for New York, based on the Occupational Employment Statistics survey by US Bureau Labor Statistics.[51] For most of the occupations, the dataset includes the average annual wages while for hourly-paid occupations like actors the dataset includes the mean hourly wages. For both cases, we estimated the average daily wage per occupation.

## Construction of contact network

We use a degree preserving configuration model to construct the contact network as follows.

1. We define N=200,000 representative workers each with a specified O*NET occupation as nodes in our contact network. The proportion of workers in each occupation matches workforce data for New York City. The total number of workers is the minimum number that allows for at least one worker of each occupation.
2. Each worker is assigned a contact degree for home, work and transport. The work contact degree is determined by the proximity score for the worker occupation, home and transport contact degrees are drawn from fixed distributions independent of occupation (see SI).
3. The nodes are connected to form the contact network using the configuration model[52] (see Figure 2).

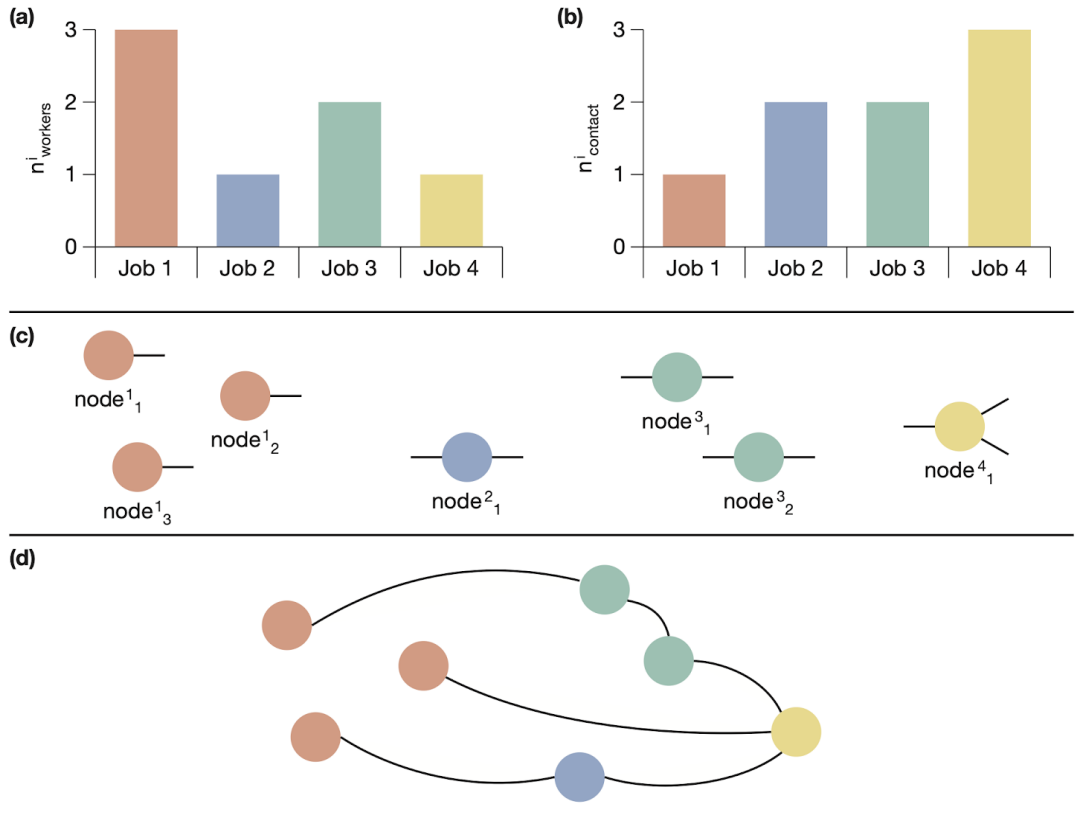

**Figure 2: Assignment of work contact links to nodes representing workers and construction of contact network using configuration model procedure. Clockwise from top left (a) worker nodes are created according to the empirical distribution in the workforce (b) work contacts are assigned according to the occupation wise proximity index (c) half-links are attached to nodes according to work contact degree (d) pairs of half-links are joined at random to form a full network.**

Each node has as many half links as the number of home, work and transportation contacts of the worker represented by the node. Pairs of half links for each category of contacts are then connected randomly until all half-links are paired up generating the final contact network.

## Epidemic simulation

Our simulation of worker based epidemic spread and containment proceeds as follows.

1. We construct an SEIR model[53] in which each worker may be in one of the four epidemic compartments. At the initial time step all nodes are considered as susceptible. An infection is seeded with one randomly chosen infected node.
2. Once the epidemic has reached 200 nodes (i.e. 0·1% of the synthetic population), we implement one of the intervention strategies described below.
3. For each epidemic simulation, we calculate the severity of the epidemic and the economic cost of furloughing workers and workers who are unproductive due to illness.

## Epidemic parameters

We emphasise that our methodology is designed for infectious disease epidemics in general. In this case we parameterise our SEIR simulations using the epidemic parameters within the reported ranges for the COVID-19 virus.

We make use of a standard SEIR model in which each worker is assigned to an epidemic state. Specifically, during the epidemic process, each node is in one of four states: susceptible (S), exposed (E), infected (I), and recovered (R). A susceptible node will become exposed for a certain period after being infected, and will transit into the infected state with a certain probability. At each time step, an infected node can recover with a fixed recovery rate. Both exposed and infected nodes can transmit the disease to a susceptible neighbor in the network with the same infection rate.

With a fixed $R_0$, we derive an estimated transmission rate per contact

$$1 - (1 - (\frac{R_o}{<d>})^{(\frac{1}{(14 + 5.1)})})$$

with the mean-field approximation assuming that each node has the same connectedness equal to the mean degree and recovers in <t_recovery> + <t_incubation> days if it is infected.

| $<R_0>$ | 2·5[54,55,56] |
| --- | --- |
| <t_recovery> | 14 days[57] |
| <t_incubation> | 5·1 days[58,59,60] |

Each epidemic simulation is seeded with one randomly infected node. Once the epidemic grows to more than 200 infected nodes, we implement a policy intervention into the network. If the epidemic dies out before reaching 200 nodes then the simulation is discarded, as we are here interested in modelling only those critical scenarios where human or policy interventions may be important factors in affecting epidemic outcomes.

## Intervention Strategies

We seek to find the dually optimal strategy -- both in infection control and economic terms -- that a policy maker might use to contain the epidemic. This requires leveraging occupation-specific information to selectively remove worker nodes from the contact network, i.e. furloughing workers. This must be done in such a way that both the severity of the epidemic, as measured by the peak of the number of infected workers, and the economic cost of furloughing workers is jointly minimised.

| Strategy Number | Name | Description | Note |
|---|---|---|---|
| -1 | Baseline | No workers are furloughed. Epidemic proceeds uninhibited. | |
| 0 | Work From Home | All workers who can work from home have work and transport links cut. | 40% of workers are able to WFH |
| 1 | Random | In addition to those who can work from home (strategy 0), send home n% of remaining workers at random. | |
| 2 | Most connected | Additionally to #0, send home n% of remaining workers ordered by contact degree. | |
| 3 | Least Essential | Additionally to #0, send home n% of remaining workers ordered by 'essentialness'. | |
| 4 | Cheapest | Additionally to #0, send home n% of remaining workers ordered by increasing wage. | |
| 5 | Centrality | Additionally to #0, send home n% of remaining workers ordered by network centrality. | We consider various centrality metrics (see SI) and report the best performing HDA |
| 6 | Control | Additionally to #0, send home n% of nodes ordered by degree, prioritising control nodes. | We consider the Switchboard model.[61,62] We find 47·9% of nodes are identified as control nodes |

## Economic productivity

Approximation of the economic impact of a worker unable to work either due to infection or furlough is a challenging measurement task.[63] We investigated two metrics: (i) the average wage of the furloughed/infected worker and (ii) the contribution to macro-economic productivity derived from the occupational share of the larger industry-level productivity. Due to the coarseness of the best available data (and the subsequent risk of substantial measurement error) for (ii) (see SI) we proceed with measure (i).

The total cost of lost productivity of infected workers is calculated as the sum of the daily wages over the period that each worker is infected (the daily wage of each worker is based on his/her occupation). The total cost of workers who are furloughed is calculated as the sum of their daily wages over the furloughed period (the furloughed period is the difference between the length of the epidemic period and the length of the intervention period).

In the former case (wage-based calculation), the wage of a worker is considered proportional to her economic contribution and her loss of income as a loss in disposable income that can be spent to stimulate supply. We acknowledge that wage is an imperfect measure of the economic contribution of a job. However we conclude from the best quality data on industry level economic contributions that estimation is not possible (see SI). Results in the main paper correspond to (i).

# Results

Figure 3 presents a comparison between the 'no intervention' and 'furloughing 10% based on degree' strategies, respectively. Under an uninhibited outbreak (strategy -1) the peak of infected nodes occurs at a mean of 129 days and with a mean 19% of nodes infected which provides reasonable correspondence to observed outbreak dynamics. This changes to a mean of 164 days and 6% infected respectively under the strategy in which workers in occupations that can work from home have transport and work ties cut.

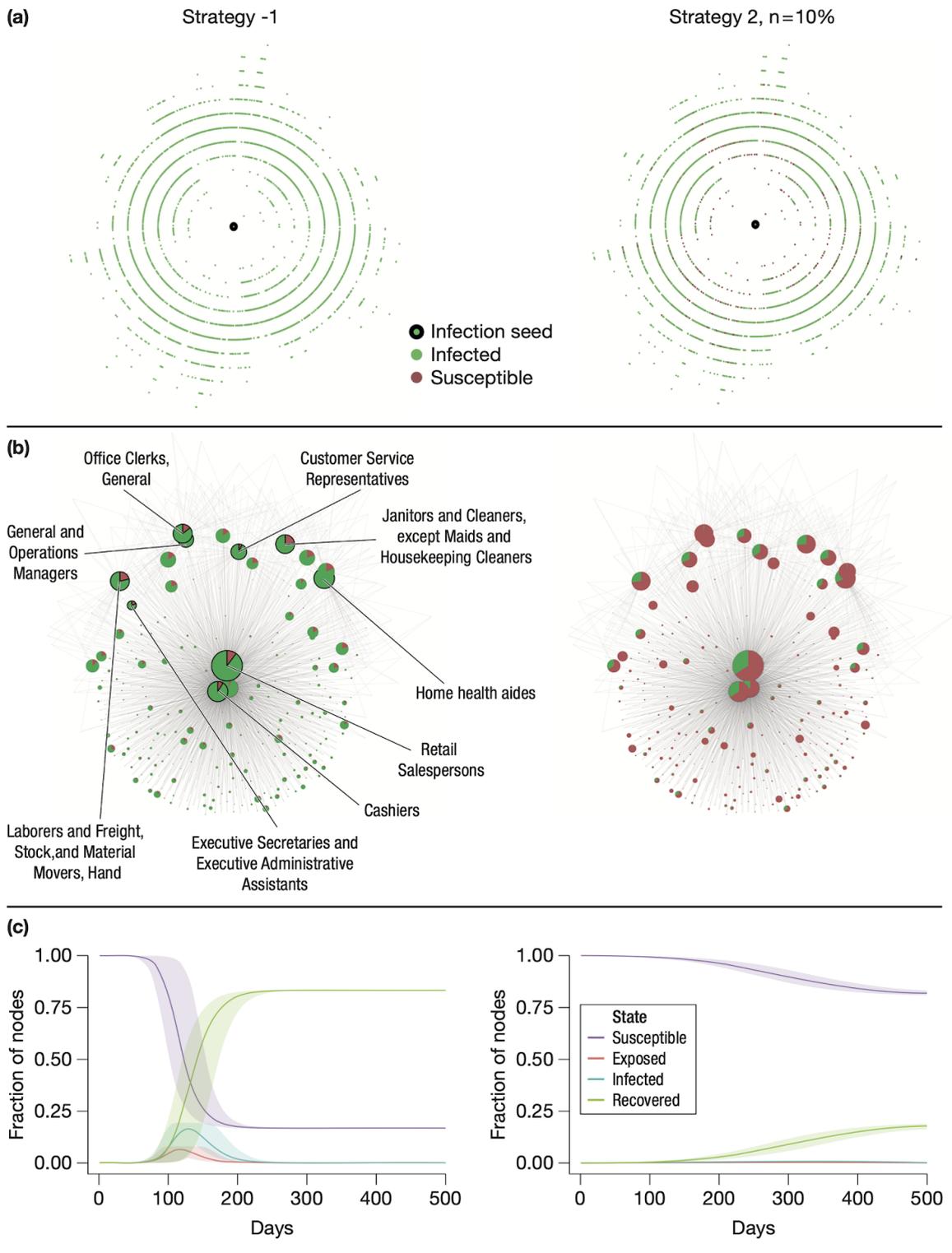

Figure 3: Epidemic dynamics under the 'no intervention' strategy, left) and the 'removing workers based on degree with n% = 10%' strategy, right). Top row (a) shows individual workers over the initial 70 time-steps of the simulations coloured by susceptible (purple) or recovered (green). Infected nodes are reduced by 19% in the second strategy. Middle row (b) shows workers aggregated into occupations

**with the size proportional to the number of workers and coloured by proportion of susceptible and recovered. Edge between two nodes indicates contacts between workers from the two occupations weighted by the number of such contacts (for visualisation purposes we only display the top three by weight edges for each node). Bottom row (c) shows the mean of 1000 epidemic curves and 95% confidence intervals.**

Figure 4 compares each strategy as a function of the stringency of the strategy (percentage of workers furloughed). In particular, the economic cost of furloughing workers calculated from wage (left) the economic cost of infected workers (middle) and the size of the peak of infected workers (right). We consider these distinct measures of epidemic severity as independent measures that policy makers would like to jointly minimise. The strategies are ordered in terms of total cost of furloughing across the range of stringency; most to least expensive, top to bottom. Generally, the measured economic loss due to infection is an order of magnitude less than the loss due to furloughing, since the infection period is ~14 days while the furlough period is the duration of the entire epidemic process (from the day that the intervention is triggered until there are no infected or exposed individuals); it is cheaper to allow a worker to be infected than to pay her wages for the duration of the epidemic.

We acknowledge that epidemic control is a careful balance of economic and human losses that is determined by some unknown and subjective multiplier of the pure cost of wages of infected workers due to secondary effects of infection e.g. lasting after effects of the disease, deaths caused by the disease, loss of quality years of life, psychological effects and so on. Likewise, furlough has negative secondary effects on individual workers.[63,64] For this reason we do not attempt to directly compare losses on these three distinct dimensions or to prescribe the extent to which workers should be furloughed. Instead we seek to compare the general behaviour of various interventions in terms of the complex dynamics of the contact network, labour market and epidemic dynamics. We consider the performance of each strategy across the entire range of severity of the strategy implementation i.e. the percentage of workers who are furloughed.

The loss due to furloughing shows a peak, at around 50% of removed nodes, for the worst performing strategies (left column rows a-c). This is related to the duration of the epidemic process (the size of the plot markers) as workers are furloughed until the epidemic process terminates. When an intermediate number of workers are removed, the epidemic is still able to spread but takes longer to fully terminate. Note that this is not attributable to herd immunity as the proportion of recovered nodes remains well below the 70-85% range. The better performing strategies (rows d-f) remove nodes more effectively, causing the epidemic to terminate sooner. In contrast to the other strategies, removing more than ~30% of nodes based on controllability, centrality and degree strategies (d-f) does not cause the epidemic period to increase significantly. This leads to a roughly linear increase in furlough costs as more workers are furloughed for the comparable duration.

In contrast, the loss due to infected workers (middle column) decreases monotonically; the total number of infected workers decreases as network edges are cut. Likewise the size of the peak of the infection (right column), representing the strain on health facilities, decreases monotonically as the 'curve is flattened'.

It should be noted that the varying behaviour of the strategies is generally not trivially attributable to the breaking apart of the underlying worker network, particularly as furloughing cuts only work and transport links preserving home contacts analogously to a home quarantine (see SI). For each strategy, the size of the Largest Connected Component decreases at roughly the same rate and contains no less than 80% of nodes across all strategies and the range of severity. Our results are broadly unchanged when home links are also severed i.e. a centralised quarantine (see SI).

We find that removing workers according to essentialness underperforms even random node removal. This is consistent with only a moderate correlation (rho = 0·19, see SI) between our proximity score and essentialness score; non-essential workers do not necessarily have a high contact degree and so their removal still allows efficient epidemic spread. Likewise removing based on wage alone performs poorly, consistent with a weak correlation between wage and proximity (rho = 0·05, see SI). For example, construction or mechanical occupations have low proximity to others. As a result, the epidemic can spread quite effectively until a large proportion of workers are furloughed (n% ~ 100%).

Removing workers based on contact degree shows a drastic increase in performance (38·6% of economic cost of the worst performing strategy; removing nodes according to essentialness) when considering the cost of infected workers. Notably, the best performing centrality metric is almost indistinguishable from node degree (38·6% vs 38·6%). Despite centrality metrics incorporating full knowledge of the network structure, they were unable to halt the spread of the epidemic markedly more effectively than simple knowledge of a workers local environment. Likewise for the controllability based strategy (38·6% vs 38·0%). Figure 5 evaluates the overall performance of each strategy across the range of stringency (n%). We repeat our analysis using an alternate random network instantiation created using the same methodology as described above. We find our results are robust to these alternative specifications (see SI).

Control nodes of a network are defined as the nodes which, if their state can be manipulated, allows for the state of the entire network to be driven to a desired configuration. Under the model of switchboard controllability (considered the most appropriate model for epidemics), we find that 47·9% of nodes in our model are identified as control nodes. Such a large proportion implies that the contact degree has a low capacity to drive controllability, which is consistent with previous studies of social networks[65] in comparison to more explicitly hierarchical networks such as those belonging to organisations or neuronal structures. Consequently, we find that a controllability based strategy offers no significant improvements in performance over a degree based strategy. This is despite incorporating full knowledge of the network structure.

These empirical results illustrate the difficulty of fully controlling an epidemic ex-ante, even with the advantage of full information of the contact network and justify the use of manual[37,56,66] and digital contact tracing[67,68] for containment as a local and dynamic strategy (as opposed to static and global one). Correspondingly, we find that removing control nodes preferentially is not a strategy that performs well, which suggests that the dynamics of epidemics might not amenable to network controllability.[65]

Given that our contact network is not empirical, it is possible that our results arise as an artifact of how the network was constructed from occupational characteristics. We emphasise that to our knowledge there does not exist a publicly available contact network larger than ~$10^2$ nodes, and none of any size that considers occupational subpopulations, that could be employed directly or bootstrapped against. Nevertheless we compare our generated network to a number of smaller empirical networks from different contexts including workplaces, schools and conferences (see SI). We find congruence between most metrics excluding those that were not explicitly coded in our configuration model such as transitivity and assortativity. This includes the high density of control nodes, in the range 40-50%.

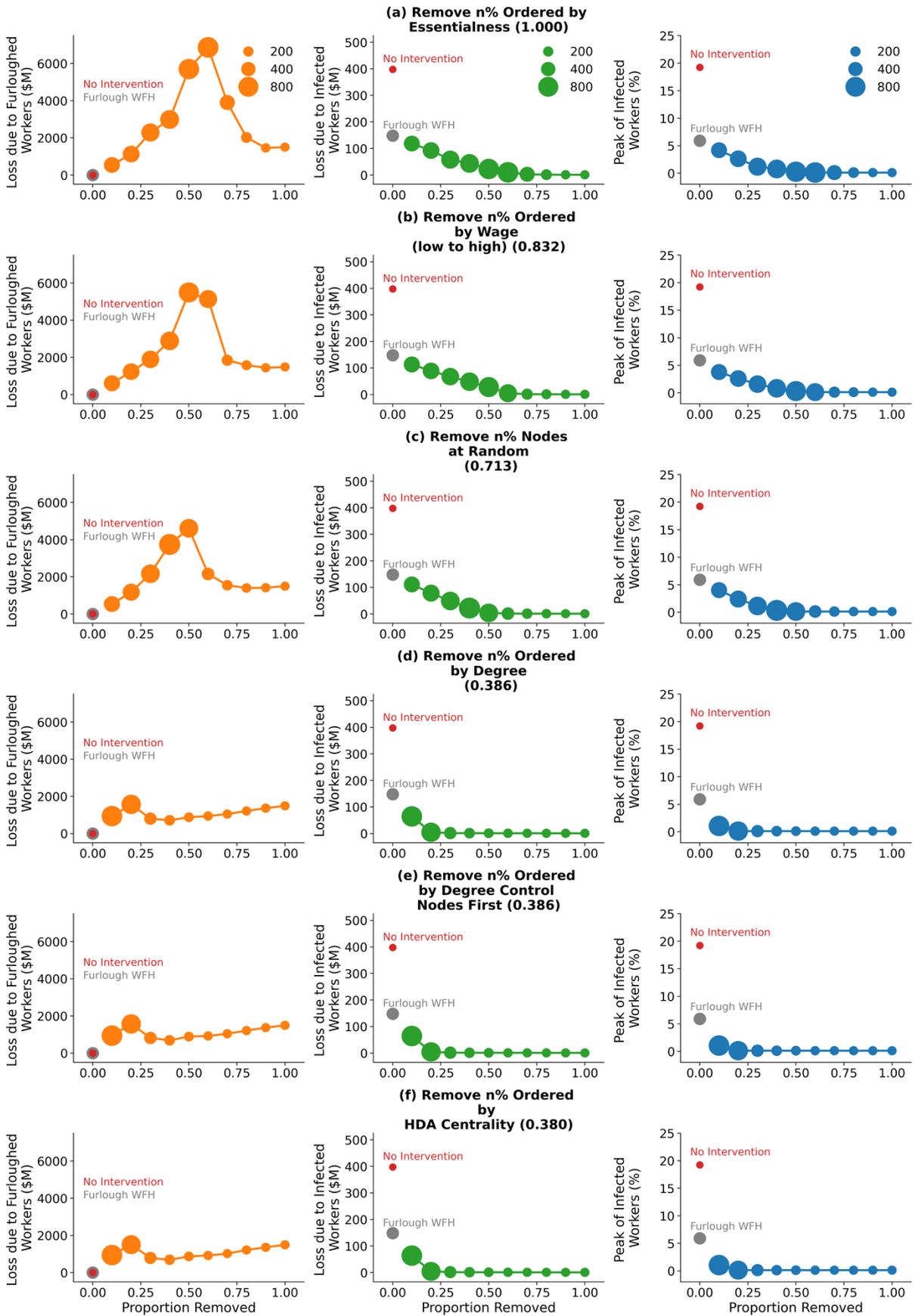

**Figure 4: Comparison of occupation furloughing strategies ordered by economic cost of furloughing (the cost is indicated in brackets next to the strategy**

description relative to the worst performing strategy). The total wage loss due to furloughing (left column), wage lost due to infection (middle column) and proportion of workers infected at peak of epidemic curve (right column) as a function of the proportion of workers removed. The fixed points for null strategies of no intervention (red point) and furloughing only workers who can work from home (grey point) are marked for comparison. Markers are sized by the duration of the epidemic period in days, defined as the time for the number of infected and exposed workers to drop to zero. Each point is an average over 1000 epidemic simulations.

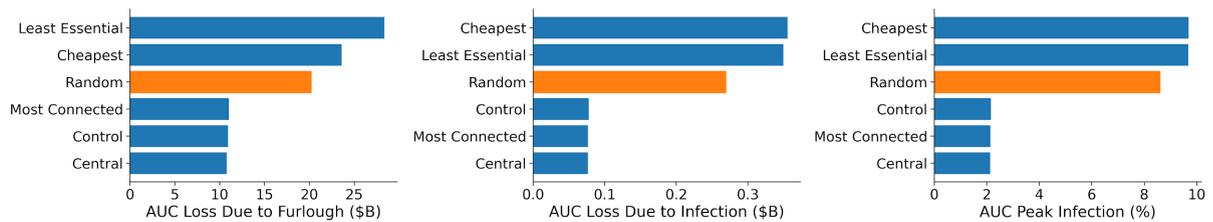

Figure 5: The aggregated performance of each strategy across the full range of severity of implementation for each measure; cost of furloughing workers (left), cost of infected workers (middle) and peak of infection (right). Error bars representing 95% confidence intervals are no greater than 0·7% of the mean value and so are omitted here for clarity.

# Discussion

In this paper we have described a method to construct epidemic contact matrices between detailed occupational sub-populations using publicly available data. We use these contact matrices as input to SEIR simulations in an urban environment and compare various epidemic containment strategies in terms of their economic and human cost. We emphasise that our goal is not to predict the spread of COVID-19, but rather to compare various occupation based containment strategies more generally using the COVID-19 epidemic as a motivation. We find that the heuristic of worker node removal according to network degree (the number of physical contacts that a worker has) performs approximately the same as more complex metrics based on complete network structure or other occupational characteristics. More broadly we note that epidemic contact networks exhibit low levels of controllability.

Our findings demonstrate that the structure of the contact network heavily influences disease dynamics in non-trivial ways. For example, furloughing a small proportion of workers can lead to pruning of the network in such a way that the epidemic persists for a long time, albeit at low levels, leading to a long and costly furlough. Intuitive strategies such as furloughing workers based on the essentialness of their job, by wage or at random all perform poorly on this basis. In contrast, network-based metrics such as degree and centrality are able to reduce the peak of the infection (flattening the curve) and also reduce the epidemic period.

Our methods exist in a relative paucity of sufficiently rich data to simulate real-world dynamics with high fidelity. Empirical contact networks, whether self-reported or passively sensed, are only freely available on a small scale. Networks with any demographic information are rarer still, while contact networks with occupational information are not to our knowledge available at all. In a workplace the correspondence between physical contact and social ties is likely to be weaker as some jobs, particularly service jobs, require a great deal of face to face contact. Consequently social ties are likely not to be a trivial proxy for physical contact ties, as observed previously.[35] Thus social networks are not a viable proxy for worker contact networks.

While we have successfully attributed much of our results to node degree, we expect that our results would differ if our method were repeated in a real contact network reflecting transitivity and hierarchical organisational structure. While the role of empirical structure in network spreading is non-trivial, it is likely that the global spread would be slower[69] than observed in our degree preserving configuration model.

A more precise critique is that our configuration model connects edge stubs between nodes at random, once a degree has been assigned based on occupation. It might be expected that the edge structure in an empirical contact network due to organisational structure or other heterogeneities might lead to very different behaviour than those we have observed. Shuffling the edges of our contact networks i.e. an Erdos-Renyi network with equivalent average edge density, we confirm that considerable structure present in our degree preserving configuration model is destroyed. This can be seen by comparing the distribution of various centrality measures (see SI). Likewise our benchmark empirical contact networks exhibit comparable behaviour. The finding that the function of a network is to large extent determined by the node degree sequence, and less by the precise connections is consistent with findings on the controllability of a large corpus of networks.

An additional caveat relates to the inherent difficulty of assessing the full economic effect of workers being unable to work. While such epidemic containment policies are generally considered on the basis of cities or states, industries are dynamically linked through national economies and global supply chains. The full economic and social cost of epidemics and associated non-pharmaceutical interventions is so complex to model as to be out of scope of this single paper.

Subject to the limitations noted above, our findings may have important implications for the practical implementation of epidemic control. Public health services have a limited number of tools available for epidemic containment, primarily furloughing of workers, contact tracing of identified infected individuals and pre-emptive surveillance of citizens. The precise implementations of these strategies require careful trade-offs between economic, social and privacy costs.

Our findings make the case for better data on contact patterns broken down by occupation. However we demonstrate that a policy of furloughing workers based on total number of contacts is far more effective at minimising epidemic spread and the cost of furloughing workers as compared to other heuristics such as furloughing based upon essentialness or furloughing a proportion of the workforce at random. However we find that there are diminishing returns to increased knowledge of the contact network beyond degree. Heuristics such as centrality or control nodes are not able to significantly outperform a simple count of contact degree. This is despite a huge increase in both the amount of invasive personal data required to reconstruct a full contact network and the complexity of its collection. Practically speaking, worker degree could be straightforwardly estimated using Bluetooth proximity to other phones, or GPS co-location. Each contact need not be identified beyond a unique untraceable ID to uniquely identify repeated contacts and no further information need be shared between individual devices or with a centralised authority.

The COVID-19 epidemic has caused many profound societal changes that are unlikely to be reversed even once the disease abates due to mass vaccination. This includes scrutinisation of the nature of work in light of vast changes in demand across sectors, variable infection across occupations, the large scale adoption of remote working and challenging deeply ingrained understandings of workplaces.

The ability to successfully automate the skills of workers, often known as skills based technological change, depends not only on overcoming the underlying physical or engineering challenges. Moreover the automating technology must be economical and must be socially and politically acceptable in order to successfully replace human labour. The changes put in motion by the COVID-19 pandemic are likely to drastically alter these considerations, in some cases leading to 'automation forcing'.[10]

Presently 'goods producing' jobs that can be performed in isolation e.g. construction workers or truck drivers are considered to be at high risk of automation. In contrast 'service producing' jobs with a high degree of personal contact e.g. fitness instructors or nannies, are at relatively lower risk. Consequently an increase in employment share has been observed in the latter category.[70]

We have demonstrated that the nature and number of connections between workers; the structure of the contact network, has vastly more potential to effectively stop epidemic spread than a single attribute of each occupation such as wage or essentialness. Looking to the future, we might reasonably expect that research and development of automation technologies will refocus to target the skills present in occupations with a high degree of contact with others. This could in turn lead to a change in these prevailing dynamics within low wage jobs.

# Acknowledgements

We thank Dr. Weihua Li for contributions made to earlier stages of this project. The authors were unable to contact Dr Li to receive approval for submission of the completed manuscript. We were therefore unfortunately precluded by arXiv submission policies from including Dr Li as an author. We also acknowledge Jurgen Rossbach for help with editing figures. This research made use of the Rocket High Performance Computing service at Newcastle University

# Supplementary Information

## Occupational Proximity

We consider five relevant dimensions of proximity to others and infection risk, by occupation taken from the O*NET database.

1. Exposed to disease or infections[71]
2. Performing for or working directly with the public[72]
3. Communicating with persons outside of the organisation[73]
4. Deal with external customers[74]
5. Physical proximity[75]

Several of these dimensions include both a 'level' and 'importance'. Since these are in general almost collinear, we just consider only the level measure. The value of the level measures ranges from 0 to 100. Less than 2% of values were missing and were imputed with the mean value.

Given that O*NET does not provide an explicit measure of physical contacts, we strive to proxy the degree of contact with others from the measures listed above. It is expected that the metrics above are neither perfectly orthogonal nor able to completely determine the desired measure of physical contact, for example *communication with persons outside of the organisation* could be conducted via phone and not only in person. We therefore construct a composite measure of proximity from these five dimensions using a PCA. This is able to explain 53% of the variance.

In order to validate this approach, we compare the occupations that are assigned a high score via our method to those which have been reported to have high rates of infection.[76]

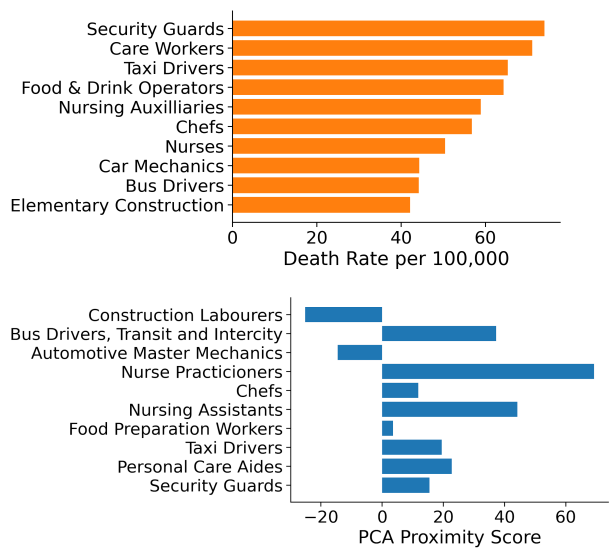

**Figure S1: The COVID-19 death rate in top ten jobs as reported in[77] (left) and the composite proximity score for the closest matching occupations in the O*NET data.**

Generally, comparing both barplots we find that similar occupations with high proximity correlates to high death rates of related jobs during the COVID-19 epidemic, although the complexity of factors that determine real death rates is way beyond physical contacts.

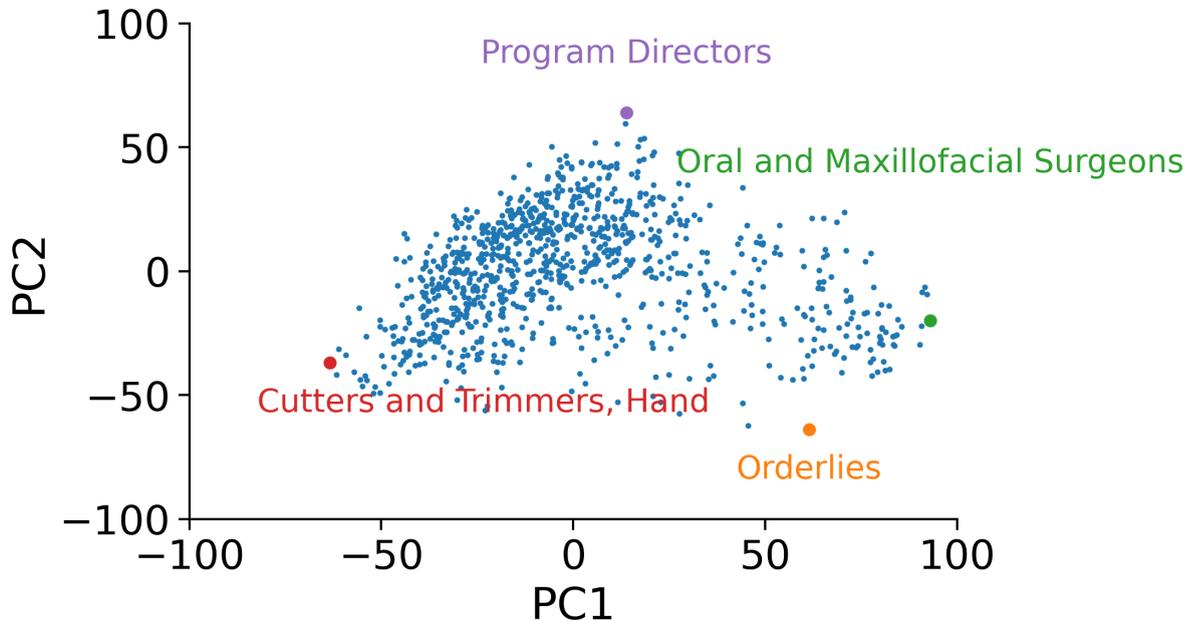

**Figure S2: Projection of selected jobs on the first and second principal components (PC).**

## Contact Degree

We derive contact degree distributions as follows. We fit log-normal distributions to several contact degree distributions from the literature.[31,34,78–81] For each, we calculate the mean and the variance and derive from these the Coefficient of Variation (CV) given as below for a log normal distribution $X$:

$$E(X) = exp(\mu + \sigma^2/2), \quad var(X) = (exp(\theta\sigma^2) - 1) * exp(2\mu + \sigma^2\theta), \quad CV(X) = \sqrt{var(X)}/E(X) = \sqrt{(exp(}$$

where $\mu$ and $\sigma^2$ is the mean and variance of the normal distribution $Y = log(X)$.

The parameters of each fit are listed in Table S1. We wish to recover the mean number of contacts of 75 as observed in[82] and to split these total contacts into work, home and transport (more accurately representing all other urban interactions: transport but also shops, cinemas, restaurants etc) while imposing a skew that is consistent with the Coefficient of Variation observed empirically. We choose a CV value of 0·7 representative of the mean and the middle of this range of CV values.

**Table S1: Parameters of lognormal distribution fitted to the degree distribution of six empirical networks.**

|  | $\mu = E(Y)$ | $\sigma = \sqrt{var(Y)}$ | $E(X)$ | $\sqrt{var(X)}$ | $CV(X)$ |
|---|---|---|---|---|---|
| Workplace | 3·55 | 0·61 | 42·05 | 28·50 | 0·68 |
| Conference | 3·62 | 0·77 | 50·37 | 45·39 | 0·90 |
| Art exhibition | 2·36 | 0·78 | 14·34 | 13·13 | 0·92 |
| Hospital | 3·27 | 0·57 | 30·91 | 19·20 | 0·62 |
| Primary school | 4·15 | 0·43 | 69·21 | 30·82 | 0·45 |
| High school | 3·07 | 0·58 | 25·51 | 15·98 | 0·63 |

**Home**: the mean home degree is derived from census information[49] which gives a mean of 2. The SD is derived from the CV=0·7. We sample from this distribution independently of occupation for each worker.

**Work**: the work degree is derived by matching the percentiles of (i) the PCA score described above and (ii) a log normal distribution with mean 36·5.[82] The SD is derived from the CV=0·7. The work degree is fixed between workers with the same occupation.

**Transport**: the mean transport degree of 36·5, is given by the remaining contact degree after accounting for home and work. Once again the SD is derived from the CV. Transport degree is sampled from this distribution for each work independent of occupation.

## Occupational Proximity to Contact Degree

As described in the main text, we have a multidimensional score of proximity in the range [0-100] for each occupation. We seek to translate that score into a work contact.

These distributions are fitted to data found in.[83] The values are then rescaled to make a mean total of 75, matching empirical data.[82]

The work degree of the node depends solely on her occupation, which is proportional to the rescaled value of its projection on the first PC of the job proximity data with the mean work degree equal to 36·5.

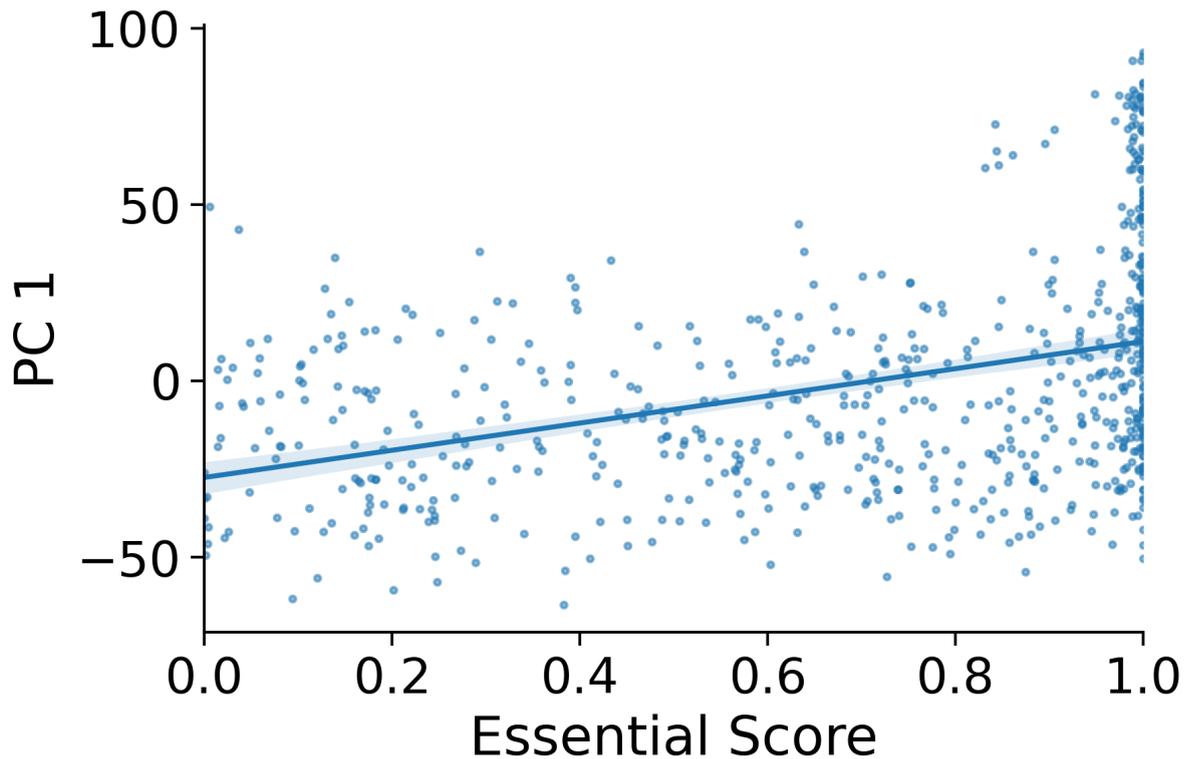

**Figure S3: Essential score from Rio-Channon et al vs first Principal Component of proximity data. Pearson correlation is 0·375.**

## Workforce Data

To generate the occupational contact network through the configuration model, we use data from the NY Workforce dataset for 2019[84] and the data for the number of contacts per occupation that we derive through the principal component analysis. The NY workforce dataset includes information about the 750 distinct occupations that exist in the NY, such as the number of employees per occupation (out of a total labour size of around 9·5M workers), their percentage in the NY workforce, and the hourly and annual average wages per worker per occupation. Each occupation in the dataset is represented by a six-digit code based on the 2010 Standard Occupational Classification System.[85] The first two digits represent the major group, the third digit represent the minor group, the fourth and fifth digits represent the broad occupation and the sixth digit represent the broader occupation. The top three occupations by number of employees in the NY workforce, are the Retail Salespersons (SOC:41-2031), the Home Health Aides (SOC:31-1011) and the Cashiers (SOC:41-2011) with 302430, 202660 and 198980 employees respectively (that is around 3·2%, 2·1% and 2·1% of the NY workforce). The bottom three occupations in the dataset are the Fallers (SOC:45-4021), Radio Operators (SOC:27-4013) and Hoist and Winch Operators (SOC:53-7041) with just 30, 40 and 40 workers respectively.

# Merging O*NET and Workforce Data

The data for the number of contacts per occupation derived by the Principal Component Analysis include information for 967 occupations. The occupations in this dataset are represented by an O*NET-SOC eight-digit code. The first six digits match the SOC coding scheme while the additional two digits indicate whether the O*NET-SOC occupation is a one-to-one match or is a detailed O*NET breakout of the SOC occupation.[86] To merge the two datasets by occupation code, we first group the occupations included in the PCA outcomes by their six first SOC digits. In that case, the number of contacts per six-digit SOC occupation is the mean of the number of contacts between the eight-digit SOC occupations that share the first six digits. After grouping the occupations (i.e. transition from the eight-digit coding system to the six-digit classification), we end up with 773 distinct occupations. 680 out of 750 SOC codes from the NY workforce dataset were matched one-to-one with SOC codes from the contacts per occupation data. We then matched 18 more occupations presented in table S2 manually. We therefore created a dataset of 678 distinct occupations having their number of employees in the NY workforce, their average wages and the number of contacts per occupation.

**Table S2: Matching unmatched SOC codes between O*NET and contacts per occupation data.**

| Occ Code in NY workforce data | Occ Title in NY workforce data | Aggregated Occ Code in contacts data | Occ Title in contacts data |
|---|---|---|---|
| 13-1020 | Buyers and Purchasing Agents | 13-1021 | Buyers and Purchasing Agents, Farm Products |
| 19-1099 | Life Scientists, All Other | 19-1011 | Animal Scientists |
| 21-1018 | Substance Abuse, Behavioral Disorder, and Mental Health Counselors | 21-1011 | Substance Abuse and Behavioral Disorder Counselors |
| 21-1019 | Counselors, All Other | 21-1014 | Mental Health Counselors |
| 29-2010 | Clinical Laboratory Technologists and Technicians | 29-2011 | Medical and Clinical Laboratory Technologists; Cytotechnologists; Histotechnologists and Histologic Technicians; Cytogenetic |

|  |  |  | Technologists |
|---|---|---|---|
| 35-2019 | Cooks, All Other | 35-2013 | Cooks, Private Household |
| 39-1010 | First-Line Supervisors of Gaming Workers | 11-9071 | Gaming Managers |
| 39-7010 | Tour and Travel Guides | 39-7011 | Tour Guides and Escorts |
| 47-3019 | Helpers, Construction Trades, All Other | 47-3016 | Helpers--Roofers |
| 49-9069 | Precision Instrument and Equipment Repairers, All Other | 49-9045 | Refractory Materials Repairers, Except Brickmasons |
| 51-2028 | Electrical, Electronic, and Electromechanical Assemblers, Except Coil Winders, Tapers, and Finishers | 51-2022 | Electrical and Electronic Equipment Assemblers |
| 51-2098 | Assemblers and Fabricators, All Other, Including Team Assemblers | 51-2093 | Timing Device Assemblers and Adjusters |
| 51-4199 | Metal Workers and Plastic Workers, All Other | 51-4192 | Layout Workers, Metal and Plastic |
| 51-7099 | Woodworkers, All Other | 51-7031 | Model Makers, Wood |
| 53-1048 | First-Line Supervisors of Transportation and Material Moving Workers, Except Aircraft Cargo Handling Supervisors | 53-1031 | First-Line Supervisors of Transportation and Material-Moving Machine and Vehicle Operators |
| 53-6099 | Transportation Workers, All Other | 53-4013 | Rail Yard Engineers, Dinkey |

| | | | Operators, and Hostlers |
| --- | --- | --- | --- |
| 29-1029 | Dentists, All Other Specialists | 29-1024 | Prosthodontists |
| 41-9099 | Sales and Related Workers, All Other | 41-9091 | Door-To-Door Sales Workers, News and Street Vendors, and Related Workers |

## Degree Preserving Configuration Model

We generate our contact network between workers using a modified configuration model as follows. We first represent the assigned numbers of home, work and transportation contacts of each worker as half-links of the nodes in the network (see Figure 2 in the main paper). We then randomly select a pair of half links for each category of contacts and connect them. The random selection is achieved by sampling nodes without replacement, in order to avoid duplications of contacts and self-loops. We repeat the process until all half-links are paired up. This technique produces a network that is not fully random as nodes are assigned a degree according to their occupation which is in turn taken from an empirical distribution of workers over occupations.

## Calculation of Transmission Rate from Reproductive Number

Let us consider how we can relate $R_0$, the mean number of new infections per infection, to $p_{inf}$ and $<d>$; the probability of infection between two nodes in a single encounter and the mean network degree. Since $R_0 \approx 2·5$ according to various studies, we would like to fix $p_{inf}$ appropriately in our simulations to be roughly consistent with this measured $R_0$.

Let us consider a 'point infection' in a single time-step (in our simulations this is a single day) in which node i interacts with its four neighbours (see Figure S4). With probability $p_{inf}$ node j is infected by node i and transitions from susceptible to infected.

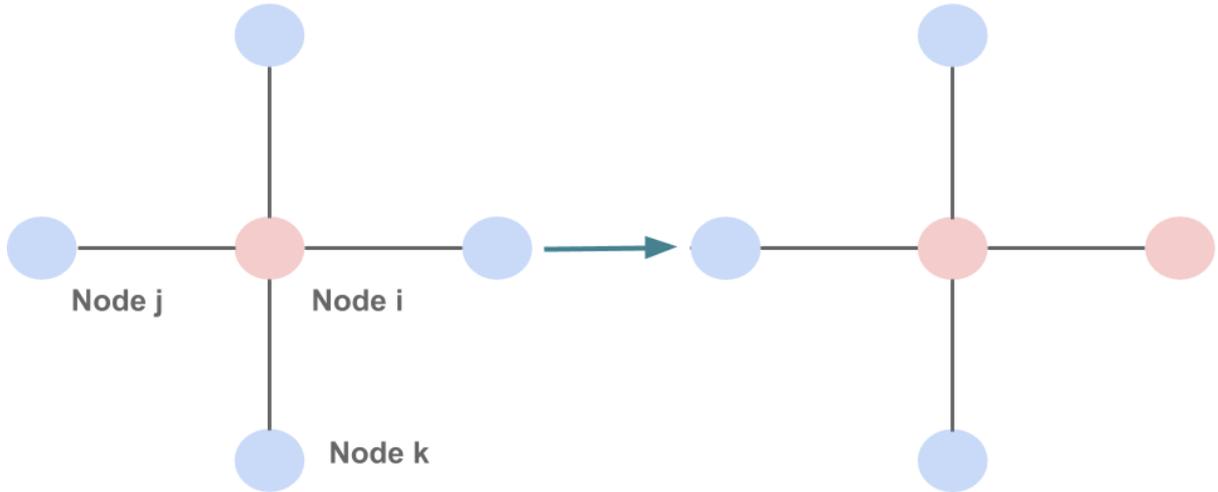

**Figure S4: A schematic illustration of a contact infection transmission**

In a single encounter the probability of j becoming infected is $p_j(S \to I) = p_{inf}$. Under n encounters, the probability of j becoming infected in at least one of the n encounters is the complement of the probability that j is not infected in any of the n encounters. The probability of not becoming infected in a single encounter is $(1 - p_{inf})$.

$$p_j^{n, p_{inf}}(S \to I) = (1 - (1 - p_{inf})^n)$$

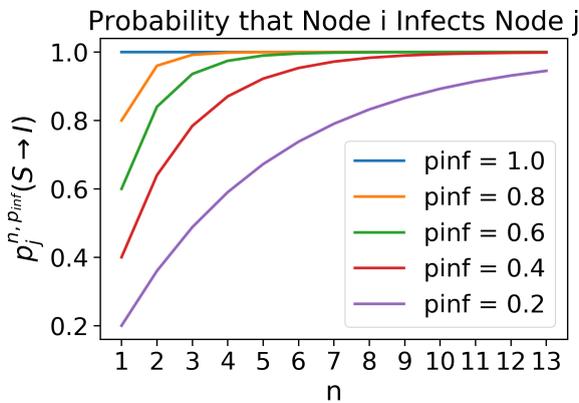

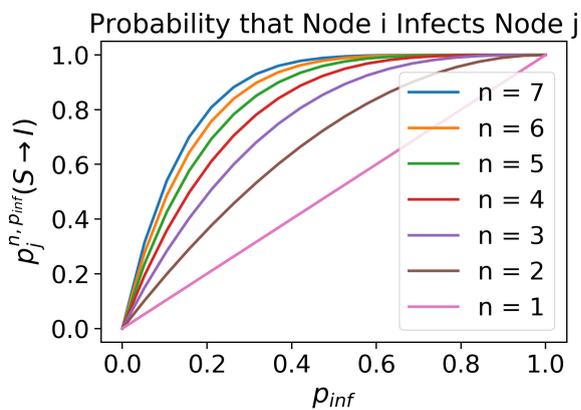

**Figure S5: The probability of a node to become infected based on the number of its infected neighbours (left) and the probability of infection in a single encounter (right).**

This can be expanded to an arbitrary number of neighbours. Assuming that there are no spillovers e.g. there is no possibility for i to infect k and then k to subsequently infect j (with i and j connected), we can calculate the expected number of new infections as the product of this probability of infection after n days and the mean degree. This is our implied $R_0$. This assumption of no second order effects is somewhat unrealistic and becomes more unrealistic as the incubation period increases.

$$R_0^{implied} = <d> \times p_j^{n,p_{inf}}(S \rightarrow I)$$
$$R_0^{implied} = <d> \times (1 - (1 - p_{inf})^n)$$

Assuming that $R_0$ is defined over the incubation period i.e. it is the mean number of new infections caused by an infected person before they are symptomatic, then we set n = 14. < d > is defined by our contact data and so $p_{inf}$ is set by our empirical $R_0$.

In fact we can calculate $p_{inf}$ directly, given our assumptions. Given our assumption of no higher order effects (i infects k who infects j) we would actually need a smaller $p_{inf}$ to recover $R_0$. So our estimated value of $p_{inf}$ is an upper bound.

$$p_{inf} = 1 - \sqrt[n]{1 - \frac{R_0}{<d>}} \sim 0 \cdot 0024$$

## Industry Based Worker Productivity

To examine the productivity of workers based upon the industry of their occupation, we employ data from the Bureau of Economic Analysis on value added by industry in producers' prices.[87] These data are calculated at the detailed industry level and exist for the years 2007 and 2012. We employ the most recent set of these data -- from 2012. We note that while industry value added has surely changed since these data were compiled, the detail industry values in BEA's estimates likely correlate decently well with actual present day value added and provide to our knowledge the highest resolution estimate of detailed industry value added in the U.S.

The BEA industry codes (NAICS) do not perfectly match the industry codes contained within the O*NET data, and so we supplement the NAICS crosswalk provided by the U.S. Census[88] with a procedure that sequentially matches industries to occupations based upon increasing levels of industry aggregation. We first match industry value added data on the full six digit NAICS codes of occupations, then match any remaining occupations and industries at the five-digit level, and so on up to the two-digit NAICS code level.

Approximately 27% of occupations have a full match, 0·3% have a five digit match, 15% have a four digit match, 28% have a three digit match, and 29% have a two digit match.

## Evaluation of Strategies

In order to evaluate and compare strategies, we consider (i) the economic loss due to furloughing of workers (ii) the economic loss due to infected workers and (iii) the size of the peak of the number of infected individuals (as a proxy for the degree of strain on health services). We wish to evaluate each strategy independently from a specific choice about the severity of the intervention i.e. the proportion of workers to furlough. We therefore calculate the areas under the curves plotted in Figure 4 of the main text; higher values are worse for all metrics. This is a simple sum given as below for strategy *s* for the furlough loss and analogously for loss due to infection and epidemic peak.

$$AUC^s_{furlough} = \sum_{n_{percent}} furloughloss^s(n_{percent})$$

Here, we present the interaction term between the cost due to furloughing and the cost of infection that complements Figure 5 in the main paper.

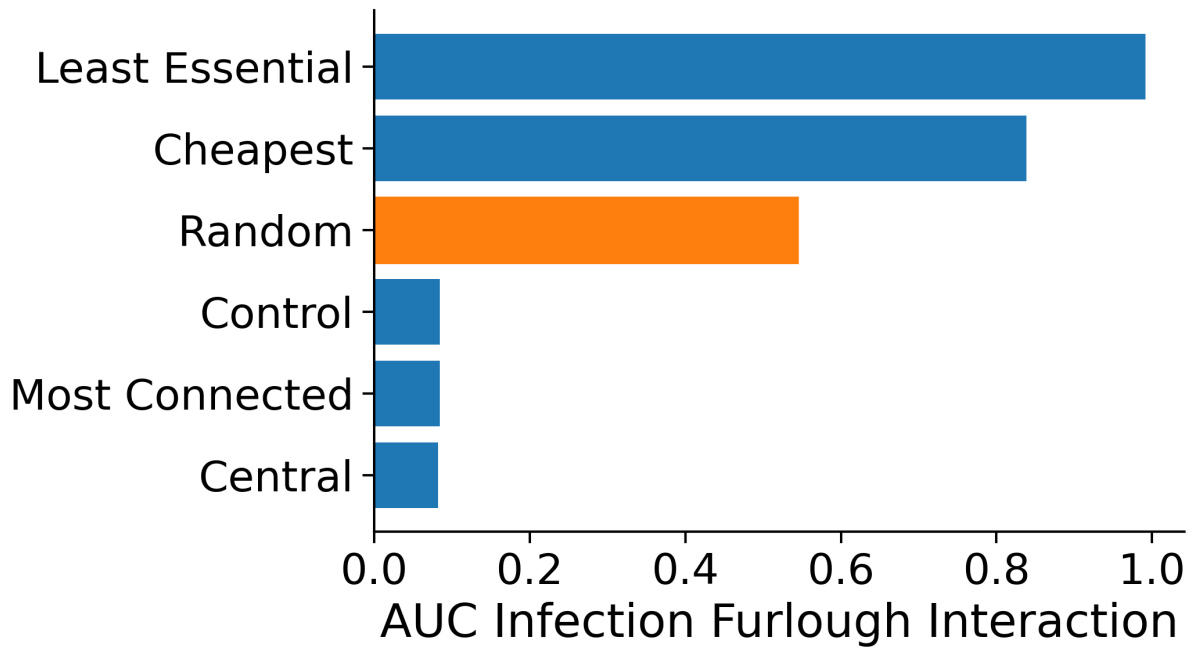

**Figure S6: The aggregated performance of each strategy presented as the interaction term between the cost due to furloughing and the cost of infection that complements Figure 5 in the main paper.**

# Centrality and Shuffled Centrality

Assigning a score to each node in a network according to its centrality is a mature area of study. We therefore compare various centrality measures, we conclude that High Degree Adaptive is the best performing and it is this metric we present in the main paper.

**Table S3: Area under curve (AUC) of cost due to furloughing and cost due to infection under strategies based on different network centrality measures. Values indicate mean ($\pm$SD) of the AUC where each point in the curve is the average of 1000 random epidemic simulations.**

| Centrality Metric | AUC Econ Loss ($B) | AUC Infected Loss ($B) |
| --- | --- | --- |
| Betweenness | 16·68 ($\pm$3·15) | 0·255 ($\pm$0·006) |
| Closeness | 11·18 ($\pm$2·34) | 0·080 ($\pm$0·004) |
| Eigenvector | 11·15 ($\pm$2·36) | 0·079 ($\pm$0·004) |
| Page Rank | 10·91 ($\pm$2·27) | 0·077 ($\pm$0·004) |
| HDA | 10·79 ($\pm$2·24) | 0·077 ($\pm$0·004) |

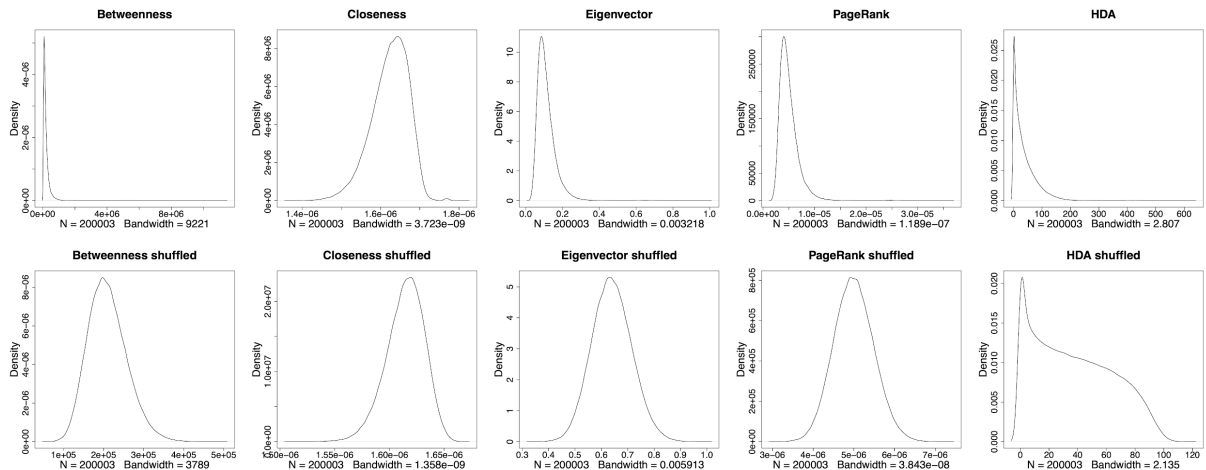

**Figure S7: Distribution (kernel density) of centrality values for our contact network constructed using the modified configuration model (top row) and under edge shuffling (i.e. an Erdos Renyi network with equivalent size and edge density).**

# Decomposition of the network in subcomponents

Our initial network is composed of a single component with 200,003 nodes. When removing the work and transportation links of nodes that can work from home the network decomposes into 3,846 components where from those, one is the giant component with 194,276 nodes and most of the other components are isolated nodes reflecting the workers that can work from home and do not have any home links. Furloughing n% nodes additional to those that can work from home, the network further decomposes in more components and the size of the giant component reduces as shown in Figures S8 and S9 for the different furloughing strategies.

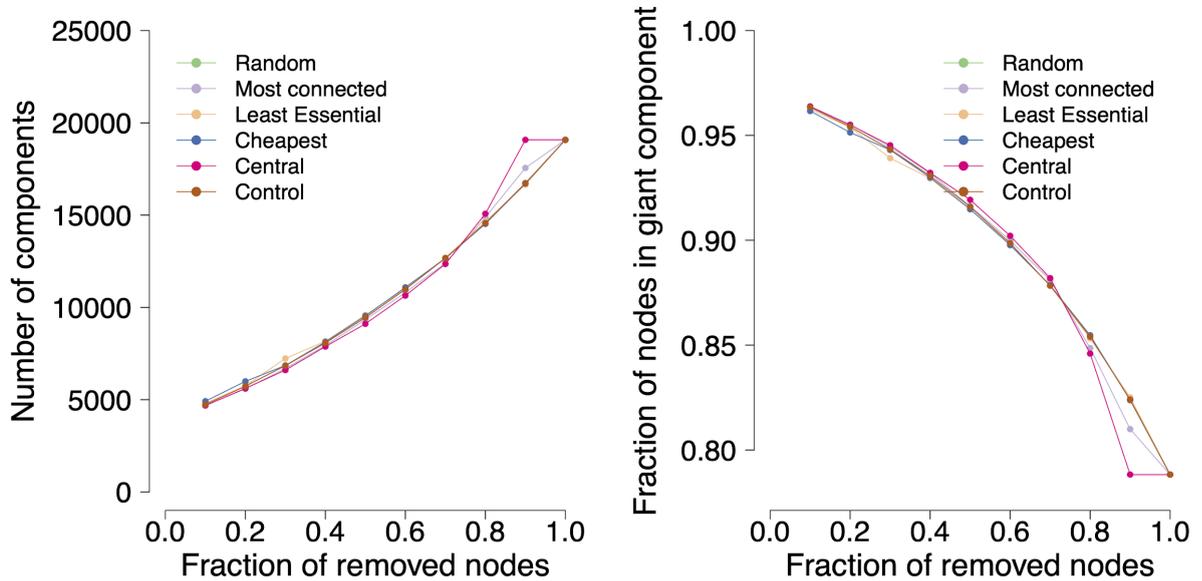

**Figure S8: Left: number of clusters in the contact network as we remove the work and transportation links of a fraction of nodes from the network. Right: the fraction of nodes in the giant component as we remove the work and transportation links of a fraction of nodes from the network.**

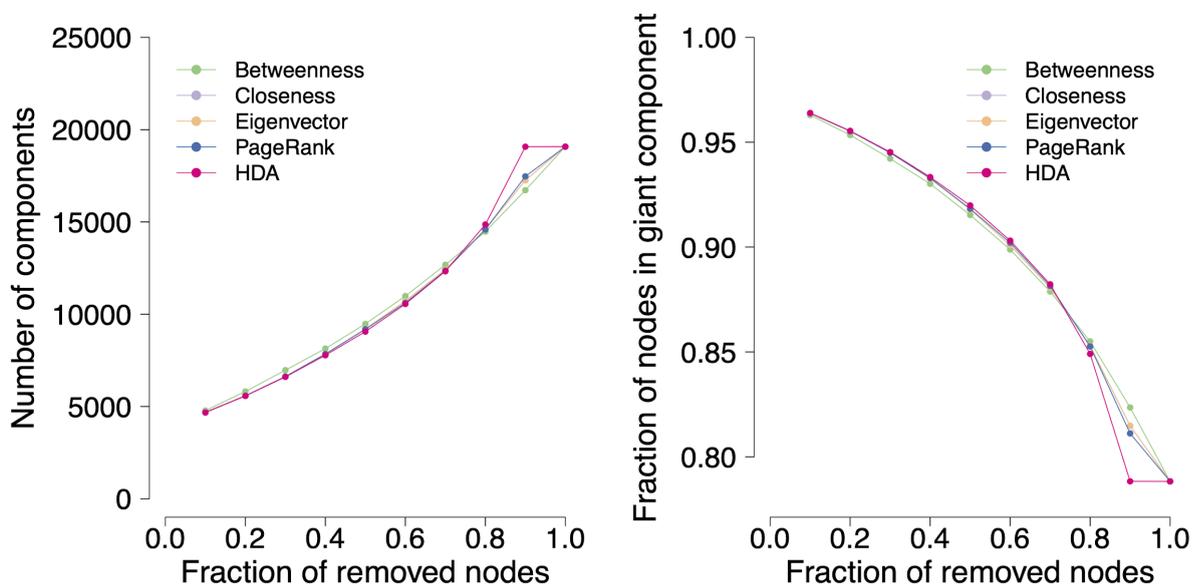

**Figure S9: Comparing various centrality measures as metrics for worker removal. Left: number of clusters in the contact network as we remove the work and transportation links of a fraction of nodes from the network. Right: the fraction of nodes in the giant component as we remove the work and transportation links of a fraction of nodes from the network.**

## Proximity and Other Worker Measures

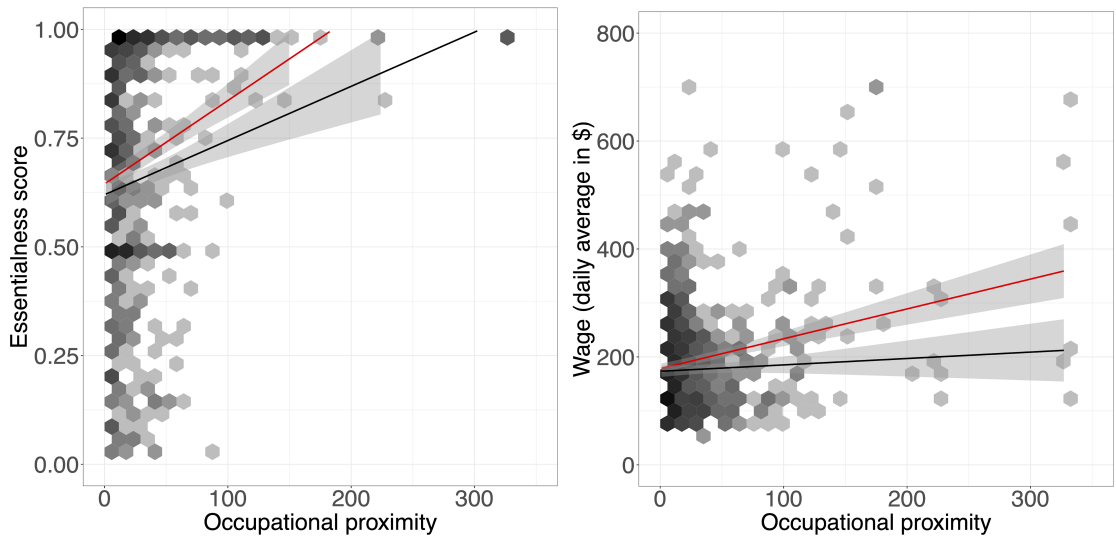

**Figure S10: Hexbin plots of proximity against essentialness score (left) and wage (right) per occupation. Red line shows unweighted fit and black line shows fit weighted by the number of workers in each occupation. The essentialness score has Pearson correlation 0·29 and 0·19 when unweighted and weighted respectively, while the wage has 0·24 and 0·05 respectively.**

## Comparison of Synthetic Contact Network with Empirical Contact Networks

We address the concern that our results are an artefact of our contact network generation model; the degree preserving configuration model. Therefore we consider several empirical contact networks from various scenarios and compare the descriptive statistics of each with our synthetic network (table S4). The "workplace" is the contact network between 217 individuals measured in an office building in France in 2015. The "conference" network represents the face-to-face interactions of 403 participants to the 2009 SFHH conference in Nice, France. The "art exhibition" is the network of contacts collected during an art science exhibition at the Science Gallery in Dublin, Ireland in 2009. The "hospital" is the contact network between patients, between health-care workers and patients and among health-care workers during four days in a hospital ward in Lyon France. The "primary school" network includes the face-to-face interactions between 242 students and teachers in a primary school, and the "high school" is the network of contacts between students in a high school in Marseilles, France. The data for those networks were active contacts as collected during 20-second intervals generating dynamic (temporal) networks in the different locations. Here we consider their static counterparts by simplifying the networks with neglection of multiple edges or by using cumulated contacts (e.g. within a day).

**Table S4: Network characteristics for our occupational contact network and six empirical contact networks.**

| Network | Nodes | Edges | Density | Min Degree | Max Degree | Average Degree | Assortativity | Transitivity | Driver nodes Switchboard (density) | Driver nodes liu (density) |
|---|---|---|---|---|---|---|---|---|---|---|
| Synthetic NY workforce | 200003 | 7491270 | 0.0004 | 13 | 633 | 74.91 | 0.003 | 0.0005 | 95,857 (0.48) | 7,430 (0.037) |
| Workplace | 217 | 4274 | 0.182 | 1 | 84 | 39.39 | 0.044 | 0.356 | 99 (0.46) | 17 (0.08) |
| Conference | 403 | 9564 | 0.118 | 1 | 169 | 47.46 | -0.081 | 0.236 | 151 (0.37) | 33 (0.08) |
| Art exhibition | 410 | 2765 | 0.033 | 1 | 50 | 13.49 | 0.226 | 0.436 | 200 (0.49) | 73 (0.18) |
| Hospital | 75 | 1138 | 0.41 | 6 | 61 | 30.35 | -0.18 | 0.587 | 30 (0.40) | 17 (0.23) |
| Primary school | 242 | 8317 | 0.285 | 20 | 134 | 68.74 | 0.118 | 0.48 | 120 (0.50) | 8 (0.03) |
| High school | 180 | 2220 | 0.138 | 2 | 56 | 24.67 | 0.046 | 0.434 | 81 (0.45) | 17 (0.09) |

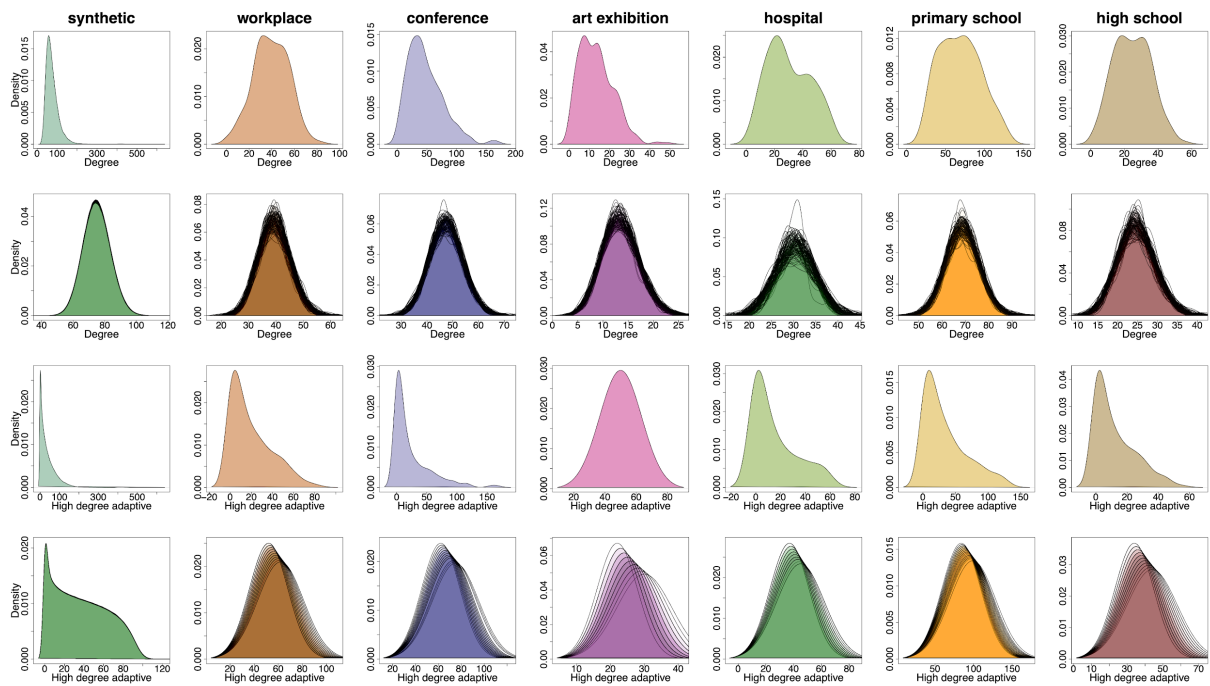

**Figure S11: Comparison of distributions (kernel densities) of observed and shuffled node degrees (first and second row respectively) and distributions of observed and shuffled high degree adaptive centrality values (third and fourth rows respectively). Left hand column is our occupational network, remaining columns are six empirical contact networks.**

# Simulations with random rewiring

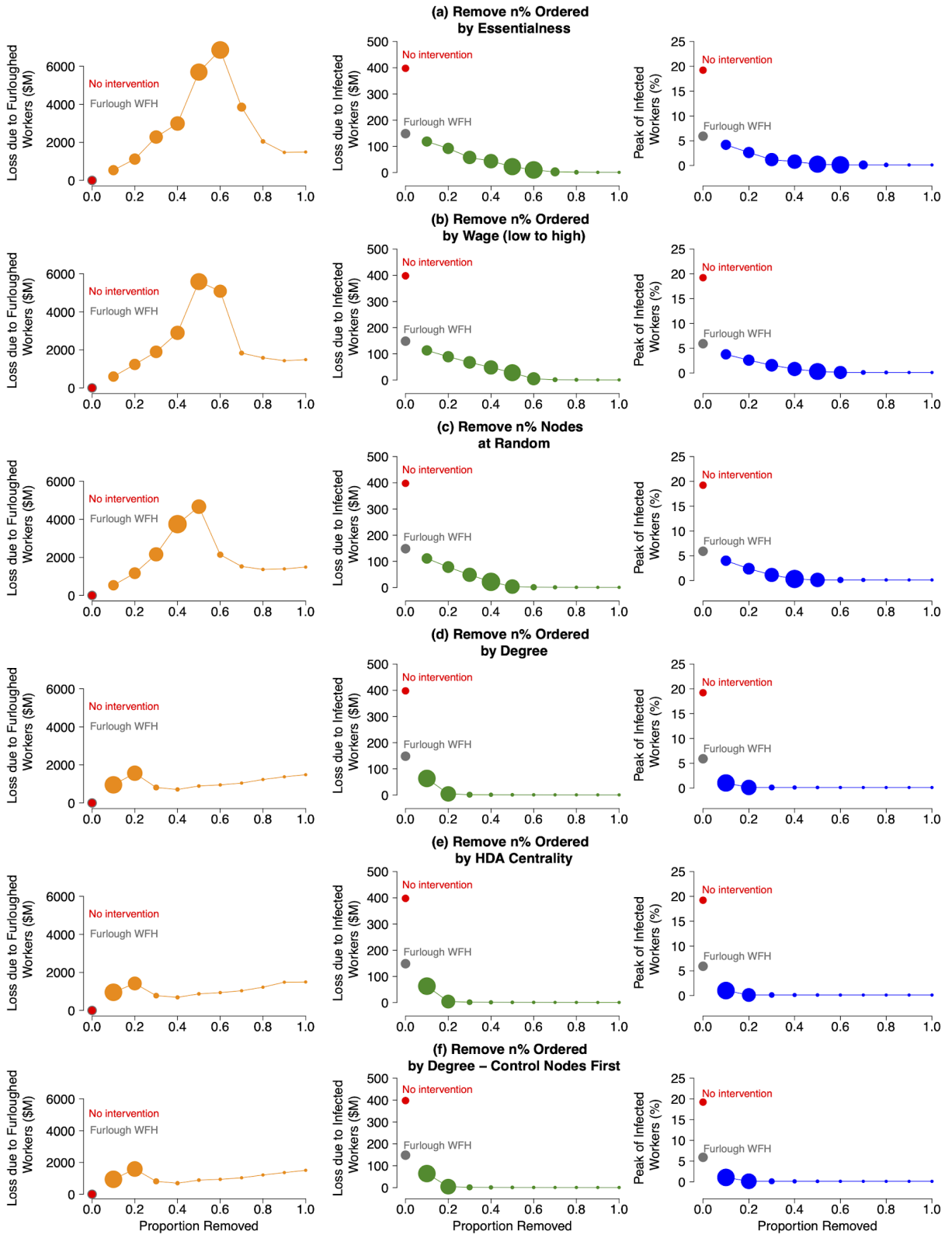

**Figure S12: This figure is similar to Figure 4 in the main paper comparing the different strategies across the full range of severities according to cost of furlough (left), cost of infection (middle) and peak of infection (right). In this case the network has 1% of links randomly rewired.**

We randomly delete 1% of home, transportation and work edges respectively and then rewire random nodes with the same proportion of links. We then repeat the simulations with the furloughing strategies on that network, and compare the outcomes (Figure S12) with the main findings shown in Figure 4 of the main paper. Each spreading scenario shown in Figure S12, is evaluated over 1000 epidemic realisations.

# Simulations with additional removal of home links of the furloughed workers

We repeat the simulations with the different furloughing strategies and we add the hypothesis of home quarantine, which means that for the furlough nodes we cut all of their links (work, transportation and home).

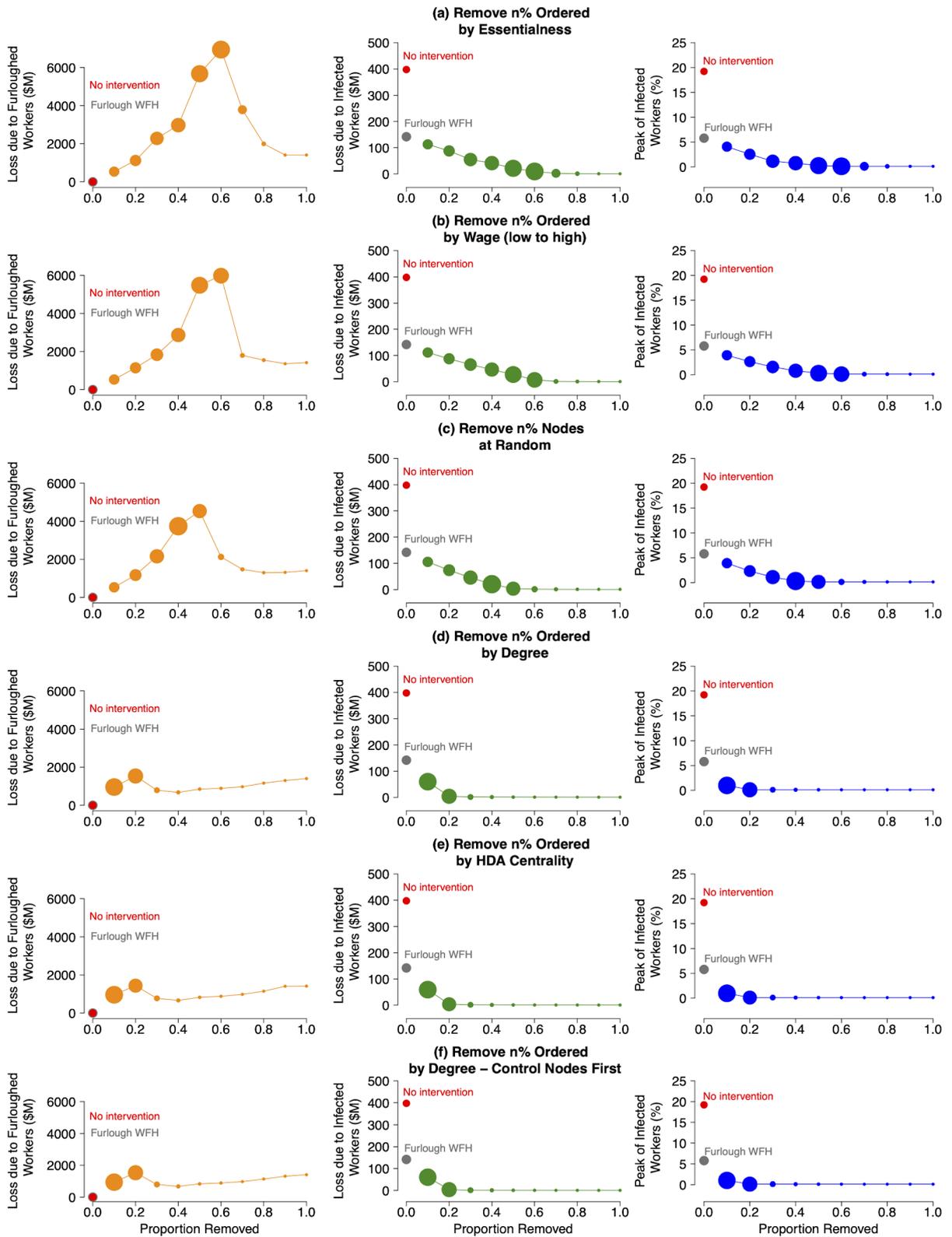

**Figure S13:** This figure is similar to Figure 4 in the main paper comparing the different strategies across the full range of severities according to cost of furlough (left), cost of infection (middle) and peak of infection (right). Here, for each strategy we remove the home links of furloughing workers in addition to the removal of their work and transportation links (i.e. a "home quarantine" scenario).

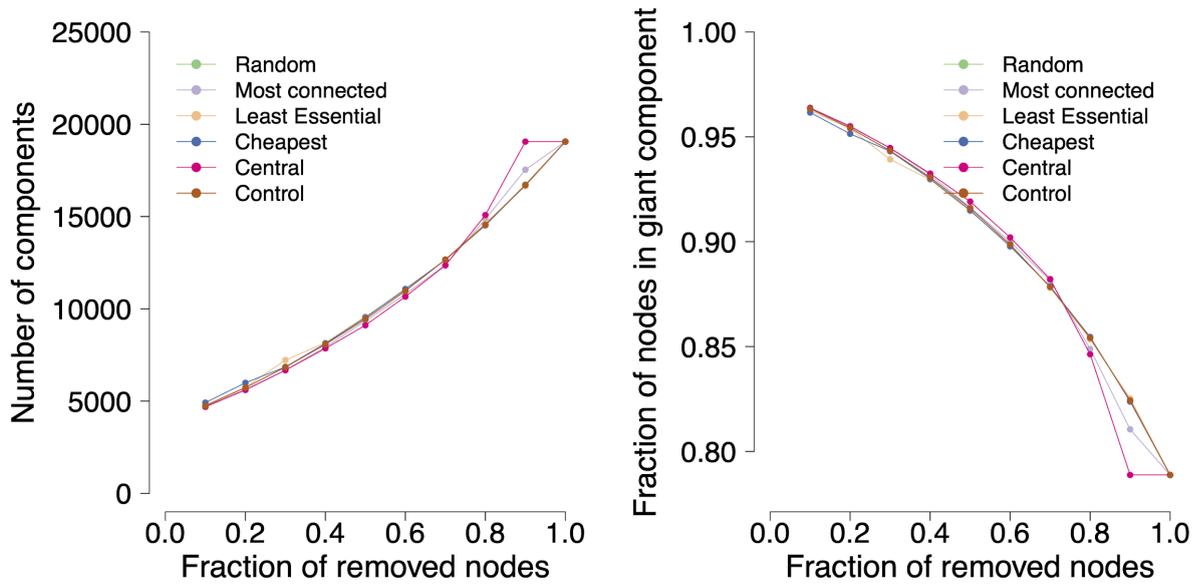

Figure S14: Left: number of clusters in the contact network as we remove a fraction of nodes from the network under the home quarantine scenarios. Right: the fraction of nodes in the giant component as we remove a fraction of nodes from the network under the home quarantine scenarios.

## Comparison with Alternate Configuration Model Instantiation

In order to ensure that our results are not an artifact of our particular random configuration of the network, we repeat our results using a different random seed.

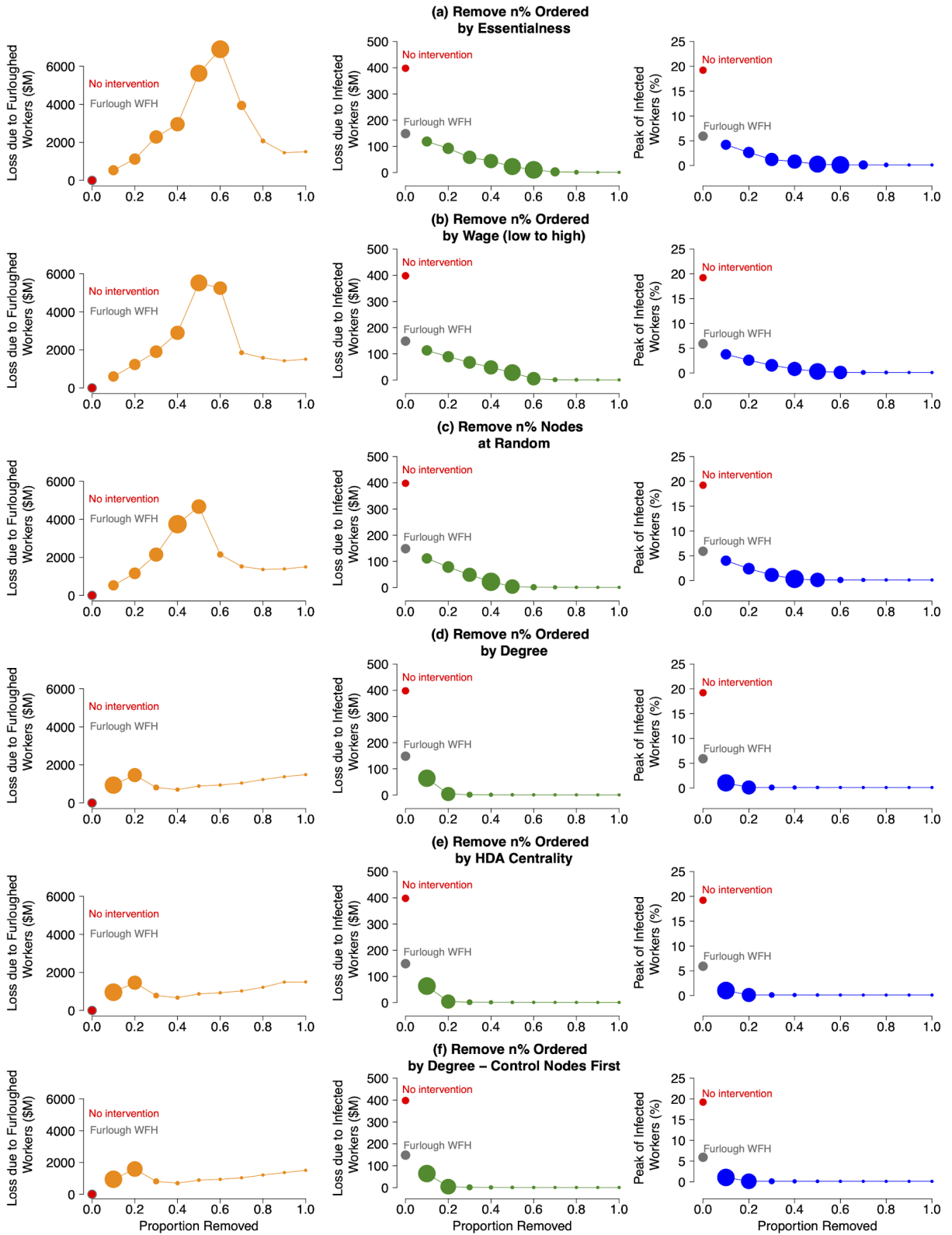

**Figure S15:** This figure is similar to Figure 4 in the main paper comparing the different strategies across the full range of severities according to cost of furlough (left), cost of infection (middle) and peak of infection (right). Here, we consider a **different network configuration drawn from the same degree preserving configuration model.**

1. Brauner JM, Mindermann S, Sharma M, Johnston D, Salvatier J, Gavenčiak T, et al. Inferring the effectiveness of government interventions against COVID-19. Science [Internet]. 2020 Dec 15 [cited 2021 Jan 19]; Available from: https://science.sciencemag.org/content/early/2020/12/15/science.abd9338.abstract

2. Glaeser E. Viewpoint: Triumph of the City [Internet]. Vol. 5, Journal of Transport and Land Use. 2012. Available from: http://dx.doi.org/10.5198/jtlu.v5i2.371

3. Bettencourt LMA, Lobo J, Helbing D, Kuhnert C, West GB. Growth, innovation, scaling, and the pace of life in cities [Internet]. Vol. 104, Proceedings of the National Academy of Sciences. 2007. p. 7301–6. Available from: http://dx.doi.org/10.1073/pnas.0610172104

4. Neffke FMH. The value of complementary co-workers. Sci Adv. 2019 Dec;5(12):eaax3370.

5. Claudel M, Massaro E, Santi P, Murray F, Ratti C. An exploration of collaborative scientific production at MIT through spatial organization and institutional affiliation. PLoS One. 2017 Jun 22;12(6):e0179334.

6. Hu B, Guo H, Zhou P, Shi Z-L. Characteristics of SARS-CoV-2 and COVID-19. Nat Rev Microbiol [Internet]. 2020 Oct 6; Available from: http://dx.doi.org/10.1038/s41579-020-00459-7

7. Zhang J, Litvinova M, Wang W, Wang Y, Deng X, Chen X, et al. Evolving epidemiology and transmission dynamics of coronavirus disease 2019 outside Hubei province, China: a descriptive and modelling study. Lancet Infect Dis. 2020 Jul;20(7):793–802.

8. [No title] [Internet]. [cited 2021 Jan 19]. Available from: https://www.ilo.org/wcmsp5/groups/public/@dgreports/@dcomm/documents/briefingnote/wcms_755910.pdf

9. The Economist. What is the economic cost of covid-19? [Internet]. The Economist; 2021 [cited 2021 Jan 18]. Available from: https://www.economist.com/finance-and-economics/2021/01/09/what-is-the-economic-cost-of-covid-19

10. Autor D, Reynolds E. The Nature of Work after the COVID Crisis: Too Few Low-Wage Jobs. 2020 Jul. Report No.: ESSAY 2020-14.

11. Kniffin KM, Narayanan J, Anseel F, Antonakis J, Ashford SP, Bakker AB, et al. COVID-19 and the workplace: Implications, issues, and insights for future research and action. Am Psychol [Internet]. 2020 Aug 10; Available from: http://dx.doi.org/10.1037/amp0000716

12. Chang S, Pierson E, Koh PW, Gerardin J, Redbird B, Grusky D, et al. Mobility network models of COVID-19 explain inequities and inform reopening. Nature. 2021 Jan;589(7840):82–7.

13. Dingel JI, Neiman B. How many jobs can be done at home? J Public Econ. 2020 Sep;189:104235.

14. Pastor-Satorras R, Castellano C, Van Mieghem P, Vespignani A. Epidemic processes in complex networks [Internet]. Vol. 87, Reviews of Modern Physics. 2015. p. 925–79. Available from: http://dx.doi.org/10.1103/revmodphys.87.925

15. Farmer JD. Economics needs to treat the economy as a complex system. In 2012.

16. Hidalgo CA, Hausmann R. The building blocks of economic complexity. Proc Natl Acad Sci U S A. 2009 Jun 30;106(26):10570–5.

17. Schweitzer F, Fagiolo G, Sornette D, Vega-Redondo F, White DR. Economic Networks: What do we Know and What do we Need to Know? Advs Complex Syst. 2009 Aug 1;12(04n05):407–22.

18. Trust and the Covid-19 Pandemic. Edelman Trust Barometer [Internet]. Spring Update. 2020; Available from: https://www.edelman.com/sites/g/files/aatuss191/files/2020-05/2020%20Edelman%20Trust%20Barometer%20Spring%20Update.pdf

19. Chowell G, Tariq A, Hyman JM. A novel sub-epidemic modeling framework for short-term forecasting epidemic waves. BMC Med. 2019 Aug 22;17(1):164.

20. Read JM, Edmunds WJ, Riley S, Lessler J, Cummings DAT. Close encounters of the infectious kind: methods to measure social mixing behaviour. Epidemiol Infect. 2012 Dec;140(12):2117–30.

21. Stopczynski A, Pentland A 'sandy', Lehmann S. How Physical Proximity Shapes Complex Social Networks. Sci Rep. 2018 Dec 7;8(1):17722.

22. Stopczynski A, Pietri R, Pentland A, Lazer D, Lehmann S. Privacy in Sensor-Driven Human Data Collection: A Guide for Practitioners [Internet]. 2014 [cited 2021 Feb 5]. Available from: http://arxiv.org/abs/1403.5299

23. Grijalva CG, Goeyvaerts N, Verastegui H, Edwards KM, Gil AI, Lanata CF, et al. A household-based study of contact networks relevant for the spread of infectious diseases in the highlands of Peru. PLoS One. 2015 Mar 3;10(3):e0118457.

24. Béraud G, Kazmercziak S, Beutels P, Levy-Bruhl D, Lenne X, Mielcarek N, et al. The French Connection: The First Large Population-Based Contact Survey in France Relevant for the Spread of Infectious Diseases [Internet]. Vol. 10, PLOS ONE. 2015. p. e0133203. Available from: http://dx.doi.org/10.1371/journal.pone.0133203

25. Ajelli M, Litvinova M. Estimating contact patterns relevant to the spread of infectious diseases in Russia [Internet]. Vol. 419, Journal of Theoretical Biology. 2017. p. 1–7. Available from: http://dx.doi.org/10.1016/j.jtbi.2017.01.041

26. van Hoek AJ, Andrews N, Campbell H, Amirthalingam G, Edmunds WJ, Miller E. The social life of infants in the context of infectious disease transmission; social contacts and mixing patterns of the very young. PLoS One. 2013 Oct 16;8(10):e76180.

27. Sun L, Axhausen KW, Lee D-H, Huang X. Understanding metropolitan patterns of daily encounters. Proc Natl Acad Sci U S A. 2013 Aug 20;110(34):13774–9.

28. Prem K, Cook AR, Jit M. Projecting social contact matrices in 152 countries using contact surveys and demographic data. PLoS Comput Biol. 2017 Sep;13(9):e1005697.

29. Mossong J, Hens N, Jit M, Beutels P, Auranen K, Mikolajczyk R, et al. Social contacts and mixing patterns relevant to the spread of infectious diseases. PLoS Med. 2008 Mar 25;5(3):e74.

30. Salathé M, Kazandjieva M, Lee JW, Levis P, Feldman MW, Jones JH. A high-resolution human contact network for infectious disease transmission. Proc Natl Acad Sci U S A. 2010 Dec 21;107(51):22020–5.

31. Stehlé J, Voirin N, Barrat A, Cattuto C, Isella L, Pinton J-F, et al. High-Resolution Measurements of Face-to-Face Contact Patterns in a Primary School [Internet]. Vol. 6, PLoS ONE. 2011. p. e23176. Available from: http://dx.doi.org/10.1371/journal.pone.0023176

32. Aharony N, Pan W, Ip C, Khayal I, Pentland A. Social fMRI: Investigating and shaping social mechanisms in the real world [Internet]. Vol. 7, Pervasive and Mobile Computing. 2011. p. 643–59. Available from: http://dx.doi.org/10.1016/j.pmcj.2011.09.004

33. Vanhems P, Barrat A, Cattuto C, Pinton J-F, Khanafer N, Régis C, et al. Estimating potential infection transmission routes in hospital wards using wearable proximity sensors. PLoS One. 2013 Sep 11;8(9):e73970.

34. Cattuto C, Van den Broeck W, Barrat A, Colizza V, Pinton J-F, Vespignani A. Dynamics of person-to-person interactions from distributed RFID sensor networks. PLoS One. 2010 Jul 15;5(7):e11596.

35. Coviello L, Franceschetti M, García-Herranz M, Rahwan I. Predicting and containing epidemic risk using on-line friendship networks. PLoS One. 2019 May 16;14(5):e0211765.

36. Potter GE, Smieszek T, Sailer K. Modeling workplace contact networks: The effects of organizational structure, architecture, and reporting errors on epidemic predictions. Netw Sci (Camb Univ Press). 2015 Sep 1;3(3):298–325.

37. Aleta A, Martín-Corral D, Pastore Y Piontti A, Ajelli M, Litvinova M, Chinazzi M, et al. Modelling the impact of testing, contact tracing and household quarantine on second waves of COVID-19. Nat Hum Behav. 2020 Sep;4(9):964–71.

38. Block P, Hoffman M, Raabe IJ, Dowd JB, Rahal C, Kashyap R, et al. Social network-based distancing strategies to flatten the COVID-19 curve in a post-lockdown world [Internet]. Vol. 4, Nature Human Behaviour. 2020. p. 588–96. Available from: http://dx.doi.org/10.1038/s41562-020-0898-6

39. Benzell SG, Collis A, Nicolaides C. Rationing social contact during the COVID-19 pandemic: Transmission risk and social benefits of US locations. Proc Natl Acad Sci U S A. 2020 Jun 30;117(26):14642–4.

40. Oliver N, Lepri B, Sterly H, Lambiotte R, Deletaille S, De Nadai M, et al. Mobile phone data for informing public health actions across the COVID-19 pandemic life

cycle. Sci Adv. 2020 Jun;6(23):eabc0764.

41. Chinazzi M, Davis JT, Ajelli M, Gioannini C, Litvinova M, Merler S, et al. The effect of travel restrictions on the spread of the 2019 novel coronavirus (COVID-19) outbreak. Science. 2020 Apr 24;368(6489):395–400.

42. Maier BF, Brockmann D. Effective containment explains subexponential growth in recent confirmed COVID-19 cases in China. Science. 2020 May 15;368(6492):742–6.

43. Kraemer MUG, Yang C-H, Gutierrez B, Wu C-H, Klein B, Pigott DM, et al. The effect of human mobility and control measures on the COVID-19 epidemic in China. medRxiv [Internet]. 2020 Mar 6; Available from: http://dx.doi.org/10.1101/2020.03.02.20026708

44. Holtz D, Zhao M, Benzell SG, Cao CY, Rahimian MA, Yang J, et al. Interdependence and the cost of uncoordinated responses to COVID-19. Proc Natl Acad Sci U S A. 2020 Aug 18;117(33):19837–43.

45. del Rio-Chanona RM, Mealy P, Pichler A, Lafond F, Farmer JD. Supply and demand shocks in the COVID-19 pandemic: An industry and occupation perspective [Internet]. Vol. 36, Oxford Review of Economic Policy. 2020. p. S94–137. Available from: http://dx.doi.org/10.1093/oxrep/graa033

46. Acemoglu D, Chernozhukov V, Werning I, Whinston M. Optimal Targeted Lockdowns in a Multi-Group SIR Model [Internet]. 2020. Available from: http://dx.doi.org/10.3386/w27102

47. U.S. Bureau of Labor Statistics. Occupational Employment Statistics [Internet]. [cited 2020 Oct]. Available from: https://www.bls.gov/oes/current/oes_35620.htm

48. Social Contact Data [Internet]. [cited 2020 Oct]. Available from: http://www.socialcontactdata.org/data/

49. Ellen IG, O'Flaherty B. Social programs and household size: evidence from New York city [Internet]. Vol. 26, Population Research and Policy Review. 2007. p. 387–409. Available from: http://dx.doi.org/10.1007/s11113-007-9036-7

50. del Rio Chanona RM, Mealy P, Pichler A, Lafond F, Farmer JD. Supply and demand shocks in the COVID-19 pandemic: An industry and occupation perspective [Internet]. 2020 [cited 2020 Oct]. Available from: https://doi.org/10.5281/zenodo.3751068

51. Occupational employment statistics home page [Internet]. 2008 [cited 2021 Mar 19]. Available from: https://www.bls.gov/oes/home.htm

52. Newman M. Networks: An Introduction. OUP Oxford; 2010. 784 p.

53. Brauer F, van den Driessche P, Wu J. Mathematical Epidemiology. Springer Science & Business Media; 2008. 414 p.

54. Anderson RM, Heesterbeek H, Klinkenberg D, Hollingsworth TD. How will country-based mitigation measures influence the course of the COVID-19 epidemic? Lancet. 2020 Mar 21;395(10228):931–4.

55. Natsuko Imai, Anne Cori, Ilaria Dorigatti, Marc Baguelin, Christl A. Donnelly, Steven

Riley, Neil M. Ferguson. Transmissibility of 2019-nCoV [Internet]. Imperial College London, UK; 2020 Jan. Report No.: Report 3. Available from: https://www.imperial.ac.uk/mrc-global-infectious-disease-analysis/covid-19/report-3-transmissibility-of-covid-19/

56. Hellewell J, Abbott S, Gimma A, Bosse NI, Jarvis CI, Russell TW, et al. Feasibility of controlling COVID-19 outbreaks by isolation of cases and contacts. Lancet Glob Health. 2020 Apr;8(4):e488–96.

57. Report of the WHO-China Joint Mission on Coronavirus Disease 2019 (COVID-19) [Internet]. 2020 Feb [cited 2020 Oct]. Available from: https://www.who.int/docs/default-source/coronaviruse/who-china-joint-mission-on-covid-19-final-report.pdf

58. Li Q, Guan X, Wu P, Wang X, Zhou L, Tong Y, et al. Early Transmission Dynamics in Wuhan, China, of Novel Coronavirus-Infected Pneumonia. N Engl J Med. 2020 Mar 26;382(13):1199–207.

59. Lauer SA, Grantz KH, Bi Q, Jones FK, Zheng Q, Meredith HR, et al. The Incubation Period of Coronavirus Disease 2019 (COVID-19) From Publicly Reported Confirmed Cases: Estimation and Application. Ann Intern Med. 2020 May 5;172(9):577–82.

60. Linton NM, Kobayashi T, Yang Y, Hayashi K, Akhmetzhanov AR, Jung S-M, et al. Incubation Period and Other Epidemiological Characteristics of 2019 Novel Coronavirus Infections with Right Truncation: A Statistical Analysis of Publicly Available Case Data. J Clin Med Res [Internet]. 2020 Feb 17;9(2). Available from: http://dx.doi.org/10.3390/jcm9020538

61. Nepusz T, Vicsek T. Controlling edge dynamics in complex networks [Internet]. Vol. 8, Nature Physics. 2012. p. 568–73. Available from: http://dx.doi.org/10.1038/nphys2327

62. Nepusz T. netctrl: Controllability of complex networks with node and edge dynamics [Internet]. Available from: https://github.com/ntamas/netctrl

63. Baranik LE, Cheung JH, Sinclair RR, Lance CE. What Happens When Employees Are Furloughed? A Resource Loss Perspective. J Career Dev. 2019 Aug 1;46(4):381–94.

64. Lee S, Sanders RM. Fridays Are Furlough Days: The Impact of Furlough Policy and Strategies for Human Resource Management During a Severe Economic Recession. Review of Public Personnel Administration. 2013 Sep 1;33(3):299–311.

65. Liu Y-Y, Slotine J-J, Barabási A-L. Controllability of complex networks. Nature. 2011 May 12;473(7346):167–73.

66. Huerta R, Tsimring LS. Contact tracing and epidemics control in social networks. Phys Rev E Stat Nonlin Soft Matter Phys. 2002 Nov;66(5 Pt 2):056115.

67. Farrahi K, Emonet R, Cebrian M. Epidemic contact tracing via communication traces. PLoS One. 2014 May 1;9(5):e95133.

68. Cebrian M. The past, present and future of digital contact tracing [Internet]. Vol. 4, Nature Electronics. 2021. p. 2–4. Available from:

http://dx.doi.org/10.1038/s41928-020-00535-z

69. Karsai M, Kivelä M, Pan RK, Kaski K, Kertész J, Barabási A-L, et al. Small but slow world: how network topology and burstiness slow down spreading. Phys Rev E Stat Nonlin Soft Matter Phys. 2011 Feb;83(2 Pt 2):025102.

70. H David DD. The growth of low-skill service jobs and the polarization of the US labor market. Am Econ Rev. 2013;

71. Work Context: Exposed to Disease or Infections [Internet]. [cited 2020 Oct 17]. Available from: https://www.onetonline.org/find/descriptor/result/4.C.2.c.1.b?a=1

72. Work Activities: Performing for or Working Directly with the Public [Internet]. [cited 2020 Oct 17]. Available from: https://www.onetonline.org/find/descriptor/result/4.A.4.a.8

73. Work Activities: Communicating with Persons Outside Organization [Internet]. [cited 2020 Oct 17]. Available from: https://www.onetonline.org/find/descriptor/result/4.A.4.a.3

74. Work Context: Deal With External Customers [Internet]. [cited 2020 Oct 17]. Available from: https://www.onetonline.org/find/descriptor/result/4.C.1.b.1.f

75. Work Context: Physical Proximity [Internet]. [cited 2020 Oct 17]. Available from: https://www.onetonline.org/find/descriptor/result/4.C.2.a.3

76. Coronavirus: Male security guards, chefs and taxi drivers among those most likely to die with COVID-19, says ONS. Sky News [Internet]. 2020 May [cited 2020 Oct]; Available from: https://news.sky.com/story/coronavirus-male-security-guards-chefs-and-taxi-drivers-among-those-most-likely-to-die-with-covid-19-says-ons-11986382

77. Coronavirus: Security guards are most at risk of dying with COVID-19, figures show. Sky News [Internet]. 2020 Jun; Available from: https://news.sky.com/story/coronavirus-security-guards-are-most-at-risk-of-dying-with-covid-19-figures-show-12015241

78. Génois M, Barrat A. Can co-location be used as a proxy for face-to-face contacts? EPJ Data Science. 2018 May 8;7(1):11.

79. Génois M, Vestergaard CL, Fournet J, Panisson A, Bonmarin I, Barrat A. Data on face-to-face contacts in an office building suggests a low-cost vaccination strategy based on community linkers [Internet]. arXiv [physics.soc-ph]. 2014. Available from: http://arxiv.org/abs/1409.7017

80. Stehlé J, Voirin N, Barrat A, Cattuto C, Colizza V, Isella L, et al. Simulation of an SEIR infectious disease model on the dynamic contact network of conference attendees. BMC Med. 2011 Jul 19;9:87.

81. Isella L, Stehlé J, Barrat A, Cattuto C, Pinton J-F, Van den Broeck W. What's in a crowd? Analysis of face-to-face behavioral networks. J Theor Biol. 2011 Feb 21;271(1):166–80.

82. Bakker M, Berke A, Groh M, Pentland A, Moro E. Effect of social distancing

measures in the New York City metropolitan area. Boston: Massachusetts Institute of Technology. 2020;

83. SocioPatterns.org [Internet]. [cited 2021 Mar 19]. Available from: http://www.sociopatterns.org/

84. U.S. Bureau of Labor Statistics. Occupational Employment Statistics: May 2019 Metropolitan and Nonmetropolitan Area Occupational Employment and Wage Estimates [Internet]. 2019 [cited 2020 Oct]. Available from: https://www.bls.gov/oes/current/oes_35620.htm

85. Website [Internet]. [cited 2020 Oct 16]. Available from: U.S. Bureau of Labor Statistics. 2010. "2010 SOC User Guide: Standard Occupational Classification and Coding Structure." https://www.bls.gov/soc/soc_2010_class_and_coding_structure.pdf.

86. Farr JL, Tippins NT. Handbook of Employee Selection. Taylor & Francis; 2017. 1005 p.

87. U.S. Bureau of Economic Analysis. Industry Economic Accounts [Internet]. 2018 [cited 2020 Oct]. Available from: https://apps.bea.gov/industry/xls/io-annual/Use_SUT_Framework_2007_2012_DET.xlsx

88. Census US. 2017 Industry code list with crosswalk [Internet]. [cited 2020 Oct]. Available from: https://www2.census.gov/programs-surveys/demo/guidance/industry-occupation/2017-industry-code-list-with-crosswalk.xlsx
44